\documentclass{aa}

\usepackage{graphicx}
\usepackage{xcolor}%

\usepackage{txfonts}
\usepackage{hyperref}

\begin{document}

\title{A scaling relationship for non-thermal radio emission from ordered magnetospheres - II. 
Investigating the efficiency of relativistic electron production in magnetospheres of BA-type stars} 

\titlerunning{Relativistic electron production in BA-type stars}
\authorrunning{Leto et al.}

\author{P.~Leto~\inst{1},
S.~Owocki~\inst{2}, 
C.~Trigilio~\inst{1}, 
F.~Cavallaro~\inst{1},         
B.~Das~\inst{3},
M.E.~Shultz~\inst{4},
C.S.~Buemi~\inst{1}, 
G.~Umana~\inst{1}, 
L.~Fossati~\inst{5},
R.~Ignace~\inst{6}, 
J.~Krti\v{c}ka~\inst{7},   
L.M.~Oskinova~\inst{8}, 
I.~Pillitteri~\inst{9},
C.~Bordiu~\inst{1},  
F.~Bufano~\inst{1},  
L.~Cerrigone~\inst{10,11},     
A.~Ingallinera~\inst{1},      
S.~Loru~\inst{1},    
S.~Riggi~\inst{1},  
A.C.~Ruggeri~\inst{1},
A.~ud-Doula~\inst{12},
F.~Leone ~\inst{13,1}      
}

\institute{
INAF -- Osservatorio Astrofisico di Catania, Via S. Sofia 78, I-95123 Catania, Italy\\
\email{paolo.leto@inaf.it}
\and
Department of Physics and Astronomy, University of Delaware, 217 Sharp Lab, Newark, DE 19716, USA 
\and
CSIRO, Space and Astronomy, P.O. Box 1130, Bentley, WA 6102, Australia
\and
ESO -- European Organisation for Astronomical Research in the Southern Hemisphere, Casilla 19001, Santiago 19, Chile
\and
Space Research Institute, Austrian Academy of Sciences, Schmiedlstrasse 6, 8042, Graz, Austria
\and
Department of Physics \& Astronomy, East Tennessee State University, Johnson City, TN 37663, USA
\and
Department of Theoretical Physics and Astrophysics, Masaryk University, Kotl\'{a}\v{r}sk\'{a} 2, CZ-611 37 Brno, Czech Republic
\and   
Institute for Physics and Astronomy, University Potsdam, D-14476 Potsdam, Germany
\and
INAF -- Osservatorio Astronomico di Palermo, Piazza del Parlamento 1, I-90134 Palermo, Italy
\and
National Radio Astronomy Observatory, 520 Edgemont Road, Charlottesville, VA 22903, USA
\and
Joint ALMA Observatory, Alonso de C\'{o}rdova 3107, Vitacura, Santiago, 7630355 Chile
\and
Department of Physics, Penn State Scranton, Pennsylvania State University, 120 Ridge View Drive, Dunmore, PA 18512, USA
\and
Dipartimento di Fisica e Astronomia, Sezione Astrofisica, Universit\'a di Catania, Via S. Sofia 78, I-95123 Catania, Italy
}

 \abstract
{
Magnetic BA-type stars typically host dipole-like magnetospheres. When detected as non-thermal radio sources, their luminosities correlate with the stellar magnetic field and rotation speed. Rotation is crucial because the mechanism undergirding the relativistic electron production is powered by centrifugal breakouts (CBOs). Small-scale breakouts occur wherever magnetic tension does not balance centrifugal force; the resulting magnetic reconnection provides particle acceleration sites.
}
{
To investigate how physical conditions at the site of the CBOs affect the efficiency of the non-thermal electron acceleration mechanism, we broadly explore the parameter space governing radio emission by increasing the sample of radio-loud magnetic stars.
}
{
High-sensitivity VLA observations of 32 early-type magnetic stars were performed in the hope of identifying new centrifugally supported magnetospheres (CMs) and associated CBOs. We calculated radio spectra produced by the {gyro-synchrotron} emission mechanism using 3D modeling of a dipole-shaped corotating magnetosphere. We evaluated combinations of stellar parameters and equatorial thermal plasma densities. The number of relativistic electrons was constrained by the need to produce the emission level predicted by the well-known empirical scaling relationship for the radio emission from magnetic BA stars.
}
{
About half of the observed early-type magnetic stars were detected, with radio luminosities in excellent agreement with the expected values, reinforcing the robust nature of the scaling relationship for CBO-powered radio emission. Comparing the competing centrifugal and magnetic effects on plasma locked in a rigidly rotating magnetosphere, we located the site of CBOs and inferred the local plasma density. We then estimated the efficiency of the CBO-powered acceleration mechanism needed to produce enough non-thermal electrons to support the expected radio emission level.
}
{
Given a constant acceleration efficiency, relativistic electrons represent a fixed fraction of the local thermal plasma. Thus, dense magnetospheres host more energetic particles than less dense ones; consequently, with other parameters similar, they are intrinsically brighter radio sources.
}


   \maketitle
%

\section{Introduction}
\label{sec:introduction}

Some main-sequence, hot, and massive stars (BA-type) possess global magnetic fields well described by a simple tilted dipole topology, known as the oblique rotating model (ORM; \citealp{Babcock1949, Stibbs1950}). In many cases, the polar magnetic field can reach kilo-Gauss strengths \citep{Shultz2019MNRAS490}. The presence of a magnetic field can affect and confine the radiatively driven stellar wind \citep{Shore1987, Shore_Brown_Sonneborn1987}, which for {BA-type} stars is relatively weak \citep{Oskinova2011, Krticka2014}, leading to the accumulation of matter around the magnetic equator \citep{Shore1990}, expressed by the magnetically confined wind shock (MCWS) model \citep{Babel1997}. The magnetic field, with the consequent presence of a large-scale magnetosphere, may be a source of photometric and spectroscopic variability modulated by stellar rotation, which is well explained by the rigidly rotating magnetosphere (RRM) model \citep{Townsend_Owocki2005}.

The magnetospheres of the BA-type magnetic stars are also sites of plasma processes responsible for phenomena detected from X-rays to radio: incoherent non-thermal radio emission \citep{Drake1987, Linsky1992, Leone_etal1994}; at lower frequencies, coherent radio emission  \citep{Trigilio2000,Das_Shultz2025}; and X-rays of thermal \citep{Babel1997,Babel1997apj,Oskinova2011,Naze2014}; and non-thermal origin \citep{Leto2017,Leto2020, Robrade2018}. Many of such observable phenomena are the direct consequence of the presence of a relativistic electron population (power-law energy distribution) inside the stellar magnetosphere. Some kind of acceleration mechanism accelerates a fraction of the co-rotating ionized material up to relativistic speed. The energetic electrons freely moving within the magnetosphere radiate in the radio regime via the gyro-synchrotron emission mechanism. 

Following the RRM model, the radio emission is also modulated by stellar rotation \citep{Leone1991,Leone_Umana1993}. The observed rotational modulation evidences that the radio emission arises from an optically thick oblique co-rotating magnetosphere, well reproduced by a 3D-model of a dipole-shaped magnetosphere \citep{Trigilio_etal2004}. The 3D-model successfully reproduced the multi-frequency radio light curves (total intensity and circularly polarized emission) of several individually studied BA-type magnetic stars \citep{Leto2006,Leto2012,Leto2017,Leto2018,Leto2020,Leto2020b}. The calculation of the wideband radio spectrum was also performed for stars where rotationally averaged multifrequency radio measurements are available \citep{Leto2021}. 

The relativistic electrons flowing through the magnetic polar caps possibly develop an unstable energy distribution \citep{Treumann2006}. In that case, highly beamed coherent radio emission arises from the auroral rings above the magnetic poles, which explains the observed coherent pulses in a radio-lighthouse fashion \citep{Trigilio2011}. 

The consistency of repeated radio measurements performed in epochs separated by many years (i.e. \citealp{Leto2012} for the incoherent emission; \citealp{Trigilio2000} and \citealp{das_chandra2021} for the coherent emission) suggests that the physical mechanism accelerating non-thermal electrons remains steady over time. The magnetospheres of BA-type magnetic stars are ideal laboratories for studying the underlying plasma processes responsible for the acceleration of the ambient thermal plasma.

In the past, it has been assumed that the engine triggering the acceleration mechanism should be linked to the presence of {a stellar wind}. As usual, far from the star, the magnetic field becomes unable to confine and channel the ionized wind material. The location where non-thermal electrons are produced has been assumed to coincide with the magnetospheric regions where the wind pressure breaks the magnetic field lines. In the resulting current sheets, electrons can be accelerated to relativistic energies \citep{Usov_Melrose1992}. The radial distance where these regions are located is quantified by the Alfv\'en radius, which can be estimated by a relation between magnetic field strength and wind parameters (mass loss rate and terminal velocity) \citep{ud-doula2008}.

In case of BA-type magnetic stars rotating fast enough to sustain a centrifugally supported magnetosphere (CM), in which the gravitational infall of the wind material is sustained by the centrifugal force \citep{Petit2013}, the wind paradigm assumed to explain the presence of relativistic electrons inside their magnetospheres has been definitively rejected \citep{Leto2021}. The weak wind typical of {BA-type} stars is not compatible with the location where the relativistic electrons have to be injected. To reproduce the wide-band radio spectra of a small sample of BA-type stars well studied in the radio domain, relativistic electrons need to be injected at a radial distance that is systematically closer to the star compared to the Alfv\'en radius location \citep{Leto2021}. Furthermore, by analyzing a larger sample of BA-type magnetic stars, \citet{Leto2021}, hereafter Paper\, I, empirically showed that their measured spectral radio luminosity ($L_{\nu,{\text{rad}}}$) depends on a combination of stellar parameters, with the stellar rotation speed having a crucial role. The evidence of a non-random behavior of the non-thermal radio emission was confirmed by \citet{Shultz2022}, who analyzed a different sample of early-type magnetic stars. Such an empirical scaling relationship seems to hold also in cases of other types of stellar or substellar (the planet Jupiter) objects surrounded by a well-ordered magnetosphere.

\citet{Owocki2022} demonstrated that the underlying physical process that powers the acceleration mechanism for non-thermal electron production inside the magnetospheres of BA-type stars is related to the power of the centrifugal breakouts (CBOs): $L_{{\text{CBO}}}$. This confirms that plasma trapped in the magnetospheres of early-type stars beyond a certain distance can break magnetic field lines, as hypothesized in the past \citep{Havnes_Goertz1984}. The CBOs occur at the distance where the centrifugal force acting on the magnetically confined plasma becomes higher than the magnetic tension \citep{ud-doula2006}, and also explain the properties of the H${\alpha}$ emission line observed in B-type stars with CMs \citep{Shultz2020, Owocki2020}.

In this paper, we report new highly sensitive VLA radio measurements of a sample of BA-type magnetic stars that were expected to be radio loud based on the $L_{\nu,\text{rad}}$ vs  $L_{\text{CBO}}$ relationship.
In Sect.\,\ref{sec:scal_rel}, we summarize the CBO scenario that explains the homogeneous radio behavior of typical BA-type magnetic stars with CMs.
In Sect.\,\ref{sec:sample_selection}, we describe the criteria followed to select the stars that belong to the selected sample.
In Sect.\,\ref{sec:vla_obs} we describe the technical details of the VLA observations.
In Sect.\,\ref{sec:obs_results} we present the results.
In Sect.\,\ref{sec:discussion}, we perform a quantitative analysis of the radio emission from BA-type magnetospheres in the context of the CBO scenario. 
In particular, in Sect.\,\ref{sec:cbo_location}, we followed a simplified approach to estimate the radial location of the region where the CBOs occur within a rigidly rotating magnetosphere;
in Sect.\,\ref{sec:spectrum_calculation}, we calculate the theoretical gyro-synchrotron spectra of a synthetic reference star; 
in Sect.\,\ref{sec:cbo_effic}, we explore the efficiency of relativistic electron production, powered by the CBOs, as a function of the model parameters;
in Sect.\,\ref{sec:non_detections}, we look for a possible common feature among the non-detected stars.
Finally, in Sect.\,\ref{sec:conclusion} we summarize our conclusions.

\begin{figure*}[]
\centering
\includegraphics[scale=1.05]{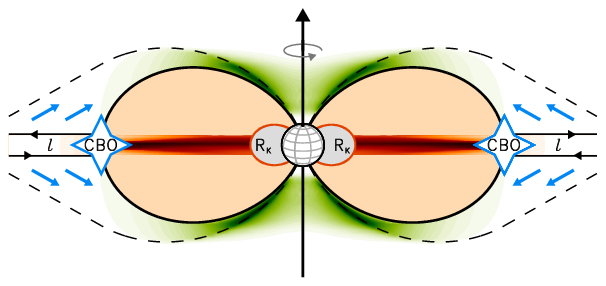}
\vspace{-5. mm}
\caption{{Cartoon summarizing} the overall scenario explaining the radio emission originating from the axisymmetric dipole-shaped centrifugal magnetosphere (CM) surrounding a typical fast-rotating BA-type magnetic star (meridian cross-section). The radio emission is a consequence of the centrifugal breakout (CBO) events suffered by the magnetically confined co-rotating plasma. In stars with CM, the CBO site is located at distances larger than the Keplerian co-rotation radius (represented by the dipole line crossing the magnetic equator at distance $R_{\text K}$). At the distance where the centrifugal force acting on the equatorial {plasma disk} (pictured by the equatorial shaded red area beyond $R_{\text K}$) wins over the magnetic tension, the thermal plasma escapes outward, and the CBOs occur, with consequent generation of an extended current sheet, where oppositely oriented magnetic field vectors exist in small-scale spatial regions. The reconnection of magnetic fields with opposite polarity is likely the acceleration mechanism responsible for isotropically distributed non-thermal electrons. The fraction of relativistic electrons (represented by the thick blue arrows) confined within the magnetic shell, defined by the last closed magnetic field line (marked by the thick black solid line) and the open field line (marked by the dashed solid line) related to the length ($l$) of the reconnection region, propagates inward along the magnetic field lines toward the poles, radiating in the radio regime via the gyro-synchrotron emission mechanism. This incoherent non-thermal emission mechanism produces a continuum radio spectrum covering a wide spectral range. The green-shaded area represents the corresponding brightness spatial distribution of the radio emission.}
\label{fig:scenario_model}
\end{figure*}

\section{The CBO scenario}
\label{sec:scal_rel}

The common behavior of the radio emission from early-type magnetic stars has been quantified by an empirical scaling relationship reported in Paper\,I, which correlates the radio spectral luminosity with the polar magnetic field strength ($B_{\text p}$), stellar radius ($R_{\ast}$), and rotation period ($P_{\text {rot}}$) as follows: 
\begin{equation}
L_{\nu,\mathrm{rad}} {\propto} B_{\text p}^2 R_{\ast}^4 / P_{\text {rot}}^2. 
\label{eq:scal_relaztion_empiric}
\end{equation}
This empirical relationship is a consequence of the physical mechanism that supports the non-thermal acceleration process, which produces the energetic electrons responsible for the magnetospheric radio emission from the stars with CMs. In a CM, the magnetically confined co-rotaing plasma extends beyond the Kepler corotation radius, that is defined by the relation: $R_{\mathrm K}=W^{-2/3} R_{\ast}$ \citep{ud-doula2006}, where $W=v_{\mathrm {rot}} / v_{\mathrm {orb}} =  \Omega R_{\ast} / \sqrt{G M_{\ast}/R_{\ast}}$ ($M_{\ast}$ stellar mass; $G$ gravitational constant; $\Omega=2\pi  / P_{\mathrm {rot}} $ angular velocity) is the dimensionless critical rotation parameter calculated as the ratio between the equatorial stellar velocity ($v_{\mathrm {rot}}=2\pi   R_{\ast} / P_{\mathrm {rot}} $) and the corresponding orbital velocity at the surface ($v_{\mathrm {orb}}=\sqrt{G M_{\ast}/R_{\ast}}$), defined by the gravitational law at the stellar equator. 

In this new vision, the stellar wind plays an indirect role in generating radio emission; it is only the thermal plasma source, with no direct role in the breaking of magnetic field lines. The wind deposits ionized matter into the stellar magnetosphere with a spatial distribution that is latitude dependent \citep{Owocki2016}. In the ideal case, when the rotational and magnetic axes are aligned, the plasma accumulation close to the magnetic equatorial plane gives rise to a {plasma disk}  \citep{ud-doula2002,ud-doula2006,ud-doula2008}. In real cases of tilted dipole magnetic configurations, the plasma is expected to be anisotropically distributed along the magnetic equator \citep{Townsend_Owocki2005}. In any case, at a distance where the magnetic tension can no longer confine the equatorial plasma, centrifugal breakouts occur. The scenario described above is pictured in Fig.\,\ref{fig:scenario_model}, which is a simplified description of a real magnetosphere that may be characterized by a warped current sheet, as evidenced by a self-consistent 3D MHD simulation of a magnetosphere with a tilted dipole \citep{ud-Doula2023}.

It was demonstrated by \cite{Owocki2022} that the power of the radio emission is directly related to the power provided by the CBOs: 
$L_{\nu,\mathrm{rad}} {\propto} L_{\mathrm{CBO}}$. 
Close to the magnetic equatorial plane, at large stellar distances, the stretching of the magnetic field provided by the centrifugal force acting on the plasma modifies the radial dependence of the magnetic field strength, making this closer to a simple radial topology that can be better described by a simple monopole ($B \propto r^{-2}$). In this case, the power released by CBOs is related to the stellar parameters (reported in cgs units) by the following relation \citep{Owocki2022}: 
\begin{equation}
L_{\mathrm{CBO}} = {B_{\mathrm \ast}^2 R_{\ast}^3} \Omega \times W \,{\mathrm{(erg\,s^{-1}).}} 
\label{eq:lrad_lcbo}
\end{equation}
At the stellar magnetic equator $B_{\ast} = B_{\mathrm p}/2$, assuming that at the stellar surface the dipole topology is not affected by the magnetospheric plasma effects. Therefore, from the Eq.\,\ref{eq:lrad_lcbo}, 
\begin{displaymath}
 L_{\nu,\mathrm{rad}} \propto L_{\mathrm{CBO}} \propto B_{\text p}^2 R_{\ast}^{4.5} M_{\ast}^{-0.5} / P_{\text {rot}}^2, 
\end{displaymath}
that is almost the same relation empirically found in Paper\,I, Eq.\,\ref{eq:scal_relaztion_empiric}, except for the weak dependence on the mass of the star.

The CBO process supports the measured spectral radio luminosity with an efficiency of the order of $10^{-19}$ \citep{Leto2022}. From a qualitative perspective, once a steady state is reached, the loss of magnetospheric plasma due to mass ejection in centrifugal breakout events is continuously replenished by wind plasma, which accumulates at low magnetic latitudes, near the magnetic equator. The CBOs occur in a well-constrained magnetospheric region, producing current sheets, where magnetic field vectors with opposite polarities exist in a small-scale region. The resulting reconnection of the magnetic fields drives the acceleration of the local electrons \citep{Hoshino2001,Zweibel2009,Dahlin2014,Ji2022} with consequent production of a relativistic electron population. These energetic electrons homogeneously fill a magnetic shell with a dipole-like topology and are responsible for the observed non-thermal incoherent radio emission produced by the gyro-synchrotron emission mechanism, which covers a wide spectral range. 

The gyro-synchrotron emission mechanism is sensitive to the local magnetic field strength, and the spatial location where the different radio frequencies originate has been shown in Paper\,I; the synthetic radio maps trace the frequency dependence of the optical depth within the stellar magnetosphere. The regions where the gyro-synchrotron arises are mainly located at the higher magnetic latitudes (shaded in green in Fig.\,\ref{fig:scenario_model}), where the magnetic topology is not expected to be strongly affected by the centrifugal stretching occurring close to the magnetic equatorial plane. In this case, the simple dipole topology is a good approximation of the true stellar magnetic field topology, confirmed by the systematic accordance between the measured multi-frequency radio light curves (of the total and the circularly polarized intensities) with those simulated using a 3D model of a co-rotating dipole-like magnetic shell \citep{Trigilio_etal2004,Leto2020b}.

\section{Sample selection}
\label{sec:sample_selection}

\begin{figure}[]
\centering
\includegraphics[width=1.9\columnwidth]{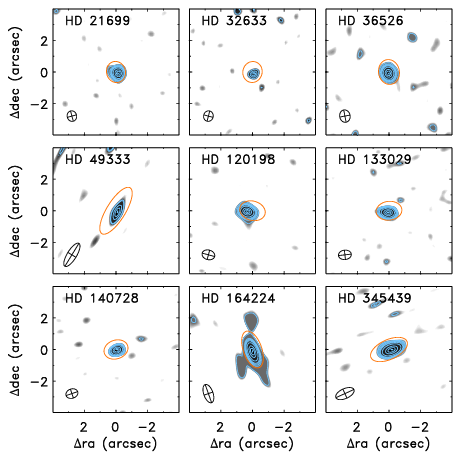}
\vspace{-5. mm}
\caption{Maps of the 9 early-type magnetic stars detected at 9 GHz (see Fig.\,\ref{fig:maps_tentativ_undetected} for the tentative detected and undetected sources). The maps are centered at the source's position corrected for the proper motion. Pixels above the $2\sigma$ threshold are displayed in grey levels (from the light grey corresponding to the $2\sigma$ level up to the black for the brightest pixel of the map). The light-blue contours of each map show the brighter pixels starting from 99\% of the source peak going down to the $3\sigma$ level. For the brightest sources, all levels spaced in steps of 99\%, 90\%, 75\%, 50\%, 35\%, 30\%, 25\%, 20\% from the peak are displayed (all levels are $\geq 3\sigma$). The red ellipses located at the positions where the stars are expected represent the sky regions covered by the First Null Beam Width (FNBW); this is about twice the Half Power Beam Width (HPBW), displayed in the bottom left corner.}
\label{fig:maps_detected}
\end{figure}

\begin{table}
\caption[ ]{Fluxes and luminosities at 9 GHz. Detections in bold.}
\vspace{-4. mm}
\label{tab:fluxes}
\footnotesize
\begin{center}
\begin{tabular}{l@{~~~~} r@{~~~~} r@{~~~~} r@{~~~~} r }
\hline   
\hline   
~~HD          &$S_{\mathrm{I}}$~~~~~~~~    &Peak            &$3\sigma$      &$L_{\nu,\mathrm{rad}}$        \\
        &{\scriptsize($\mu$Jy)}~~~~~~~  &{\scriptsize($\mu$Jy/beam)}  &{\scriptsize($\mu$Jy/beam)} &{\scriptsize(erg\,s$^{-1}$\,Hz$^{-1}$)}   \\
\hline

~~11502            & --~~~~~~~~~                 & --             &  30           &  $<1.0\times 10^{14}$            \\
~~{\bf 11503}$^{a}$   & $40  \pm  14$~~~~      & 34$^{\dag}$              &  30           &  $1.6(0.3)\times 10^{14}$         \\
~~{\bf 12767}   & $90  \pm  23$~~~~      & 50$^{\dag}$             &  33           &  $1.2(0.3)\times 10^{15}$        \\
~~15089            & --~~~~~~~~~                 & --             &  390          &  $<9.6\times 10^{14}$              \\
~~{\bf 21699}            & $255 \pm  26$~~~~       & 255            &  30           &  $9.6(0.9)\times 10^{15}$            \\
~~{\bf 32633}            & $50  \pm  12 $~~~~      & 42             &  21           &  $2.8(0.55)\times 10^{15}$        \\
~~{\bf 36526}$^{b}$            & $180 \pm  15 $~~~~      & 181            &  21           &  $3.7(0.25)\times 10^{16}$        \\
~~{\bf 49333}            & $95  \pm  24$~~~~      & 84             &  39           &  $5(1)\times 10^{15}$            \\
~~49976            & --~~~~~~~~~                 & --             &  36           &  $<4.4\times 10^{14}$            \\
~~{\bf 55522}$^{c}$   & $45  \pm  14$~~~~       & 36$^{\dag}$             &  27           &  $4.1(0.9)\times 10^{15}$            \\
~~62140            & --~~~~~~~~~                 & --             &  27           &  $<2.9\times 10^{14}$            \\
~~65339            & --~~~~~~~~~                 & --             &  36           &  $<4.1\times 10^{14}$            \\
~~74067            & --~~~~~~~~~                 & --             &  42           &  $<2.9\times 10^{14}$            \\
{\bf 103192}  & $60  \pm 23 $~~~~       & 51$^{\dag}$             &  33           &  $6.5(2)\times 10^{14}$        \\
{\bf 120198}           & $235 \pm  21$~~~~       & 200            &  24           &  $2.4(0.15)\times 10^{15}$        \\
122532           & --~~~~~~~~~                 & --             &  63           &  $<1.8\times 10^{15}$             \\
{\bf 125248}  & $40  \pm   13$~~~~      & 33$^{\dag}$             &  24           &  $3.5(0.9)\times 10^{14}$            \\
{\bf 130559}  & $70  \pm   15$~~~~      & 44$^{\dag}$             &  30           &  $5(1)\times 10^{14}$             \\
{\bf 133029}           & $175 \pm  24$~~~~       & 162            &  30           &  $5.1(0.6)\times 10^{15}$        \\
{\bf 137909}  & $90  \pm   18$~~~~      & 69$^{\dag}$             &  42           &  $1.4(0.5)\times 10^{14}$               \\
140160           & --~~~~~~~~~                 & --             &  51           &  $<2.8\times 10^{14}$             \\
{\bf 140728}           & $115 \pm  23$~~~~       & 127            &  30           &  $1.1(0.2)\times 10^{15}$        \\
{\bf 142884}  & $50  \pm  22$~~~~       & 37$^{\dag}$             &  24           &  $1.8(0.7)\times 10^{15}$        \\
145102           & --~~~~~~~~~                 & --             &  24           &  $<9.0\times 10^{14}$             \\
149822           & --~~~~~~~~~                 & --             &  24           &  $<6.1\times 10^{14}$             \\
151346           & --~~~~~~~~~                 & --             &  33           &  $<1.5\times 10^{15}$             \\
152107           & --~~~~~~~~~                 & --             &  30           &  $<1.0\times 10^{14}$             \\
{\bf 164224}$^{d}$            & $900 \pm  56$~~~~     & 918            &  39           &  $5.4(0.2)\times 10^{16}$        \\
168856           & --~~~~~~~~~                 & --             &  33           &  $<1.6\times 10^{15}$             \\
175132           & --~~~~~~~~~                 & --             &  24           &  $<2.9\times 10^{15}$             \\
183339           & --~~~~~~~~~                 & --             &  21           &  $<2.6\times 10^{15}$             \\
224801           & --~~~~~~~~~                 & --             &  21           &  $<8.0\times 10^{14}$             \\
{\bf 345439}           & $225 \pm  23$~~~~       & 221            &  39           &  $1.5(0.2)\times 10^{18}$      \\

\hline
\end{tabular}
\vspace{-2. mm}
\begin{list}{}{}
\item[Notes.] 
$^{\dag}$\,Tentative detections.
$^{a}$\,HD\,11503 detected  at 700\,MHz \citep{Shultz2022}. 
$^{b}$\,HD\,36526 detected at 890\,MHz  \citep{Das_Driessen2025}. 
$^{c}$\,HD\,55522 detected at 650\,MHz  \citep{Keszthelyi2025}. 
$^{d}$\,HD\,164224 detected at 890\,MHz and 1.4\,GHz \citep{Das_Driessen2025}.
\end{list}
\end{center}

\end{table}

\begin{figure*}[]
\centering
\includegraphics[width=1.94\columnwidth]{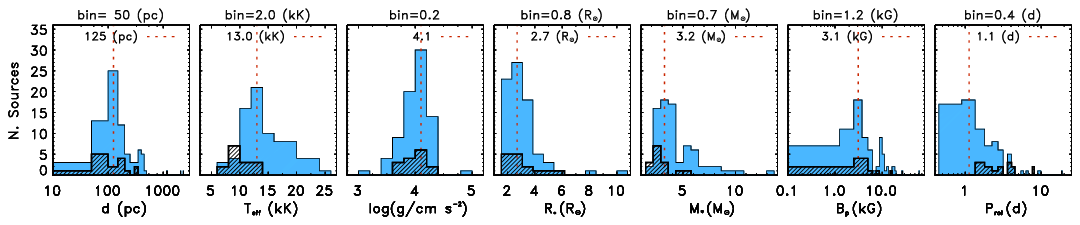}
\vspace{-2. mm}
\caption{Histograms of the stellar parameters of all the BA-type magnetic stars discovered radio loud, blue shaded area. The stellar sample analyzed in this paper (parameters listed in Table\,\ref{tab:star_param}) has been added to the large samples of stars already analyzed in other papers  \citep{Leto2021,Shultz2022,Das_Driessen2025}. The red dotted line in each panel marks the value where the distributions peak, note that the medians of the distributions ($\langle d \rangle=148$\,pc; $\langle T_{\text{eff}}\rangle=13.4$\,kK; $\langle \log g \rangle=4.05$; $\langle R_{\ast}\rangle=2.8$\,R$_{\odot}$; $\langle M_{\ast}\rangle=3.8$\,M$_{\odot}$; $\langle B_{\text p} \rangle=3.9$\,kG; $\langle P_{\text{rot}}\rangle=1.5$\,d) fall in the same bin. The parameter distributions of the non-detected stars reported in this paper are also shown (shaded area by the oblique black lines).
}
\label{fig:isto_all_param_sample}
\end{figure*}

We started from a large sample of magnetic BA-type stars having a polar magnetic field strength \citep{Sikora2019a,Shultz2019MNRAS490} at the kG level. Such a high magnetic field strength made us confident that these magnetic stars may host a centrifugally supported magnetosphere.

Then, we selected about 90 stars visible from the VLA site (source declination $>-41^{\circ}$) and for which the stellar radius, mass, and rotation period are known. The knowledge of these stellar parameters is required for the calculation of the power released by the centrifugal breakouts by using Eq.\,\ref{eq:lrad_lcbo}. 

Subsequently, we estimated the expected radio luminosity of the selected stars by using the scaling relationship between spectral radio luminosity and power released by CBOs (see Sec.\,\ref{sec:scal_rel}). The used proportionality coefficient between $L_{\nu,\text{rad}}$ and $L_{\text{CBO}}$ was then obtained by relating the two luminosities of the magnetic BA stars already analyzed in Paper\,I and in \citet{Shultz2022}, as discussed by \citet{Leto2022}.

Next, among all the magnetic BA stars selected, we restricted the sample to only those stars with a high detection expectancy. That is, we restricted the sample to stars with an upper limit of $L_{\nu,\text{rad}}$  above the corresponding radio luminosity expected by the scaling relationship, mainly drawing on the non-detections from the large sample analyzed by \citet{Shultz2022}. The upper limits of the radio luminosity were calculated via the flux detection threshold reported by \citet{Shultz2022}.  For stars not analyzed by \citet{Shultz2022}, we checked that their possible radio emission was below the VLASS \citep{Lacy2020_vlass} detection threshold and used the noise of the corresponding VLASS fields in which the star is located to calculate the upper limit of the radio luminosity.

Furthermore, based on the well-known rotational modulation of the radio emission from magnetic BA-type stars, which has been extensively studied at the radio regime, we assumed the worst-case condition corresponding to observations made when the radio emission is at a minimum. To minimize this risk, we assumed 50\% of the expected theoretical fluxes (scaling the luminosities by the Gaia distances; \citealp{Bailer-Jones2021}).

Finally, we selected targets with expected fluxes above about $30$\,$\mu$Jy. This flux limit was fixed to reach at least a $3\sigma$ detection threshold in radio maps with a noise level of about 10\,$\mu$Jy. 

Following the selection criteria described above, we collected a sample of 32 magnetic BA-type stars. The selected targets are listed in Table\,\ref{tab:star_param}, along with the corresponding stellar parameters, which for convenience we report in the Appendix\,\ref{appendix:sample_param}.

At the time when the sample of magnetic BA-type stars reported in this paper was selected, none of them had been previously detected at the radio regime. However, recent highly sensitive radio observations detected the radio emission of some of these stars \citep{Keszthelyi2025,Das_Driessen2025,Das_Shultz2025}, confirming the goodness of the selection criteria adopted.

\section{The VLA observations}
\label{sec:vla_obs}

This project is an experiment aimed at detecting the radio emission produced by the incoherent {gyro-synchrotron} emission mechanism from selected targets. In Paper\,I it was evidenced that the spectrum shape of a sub-sample of BA-type magnetic stars well studied at the radio regime is characterized by an almost flat region. We tuned to 9\,GHz the VLA observing frequency of this observing project, because the 9\,GHz radio frequency lies within the spectral range where the radio spectra are reasonably suspected to be flat. The VLA array configuration for all the observations was the B configuration. Observing parameters and other details are reported in Table~\ref{tab:vla_obs} of Appendix\,\ref{appendix:vla_obs}. 

The interferometric measurements have been processed through the VLA Calibration Pipeline using the Common Astronomy Software Applications  ({\sc casa}) package (release  6.4.1), which is designed to handle Stokes I continuum data. Images of the sky region centered at the target position have been obtained using the task {\sc tclean} (number of Taylor terms 2, image weighting "briggs", robust value 0.5).

\subsection{Detection reliability and flux measurements}
\label{sec:detect_reliability}

The coordinates corrected for the proper motions are listed in Table~\ref{tab:correct_pro_motion} reported in Appendix\,\ref{appendix:prop_motion}. We search for the radio emission inside the sky area covered by the First Null Beam Width (FNBW) and centered at the sky position corresponding to the stellar positions at the epochs of the VLA observations. The radio maps of the nine detected stars are displayed in Fig.~\ref{fig:maps_detected}. The eight targets whose radio emission has been tentatively detected are shown in Fig.~\ref{fig:maps_tentativ_undetected}, where the corresponding radio maps of the undetected stars are also shown.

The flux measurements of the detected sources and the upper limit of those targets that are below the VLA detection threshold are listed in Table~\ref{tab:fluxes}. The errors in the measured flux densities of the detected sources were computed by adding in quadrature the noise ($1\sigma$) in the maps of each source (measured in regions far from background sources), the uncertainty related to the source fitting process using a bi-dimensional Gaussian, and 5\% of the fluxes to account for calibration errors.

\section{Results}
\label{sec:obs_results}

\begin{figure}[]
\centering
\includegraphics[width=1.35\columnwidth]{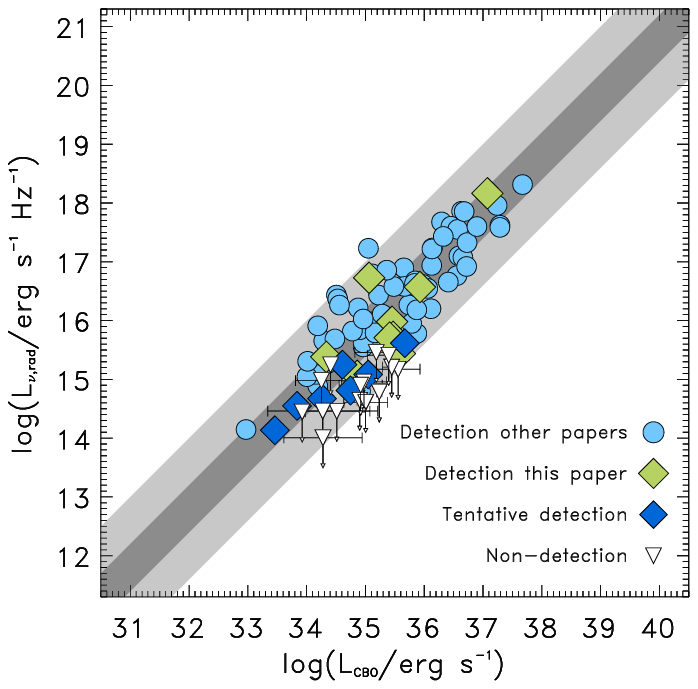}
\vspace{-5.5 mm}
\caption{
$L_{\nu, {\mathrm {rad}}}$--$L_{\mathrm{CBO}}$ diagram. Different symbols are used to distinguish the stars analyzed in this paper from the stars already analyzed in other papers \citep{Leto2021,Shultz2022,Das_Driessen2025}. The shaded diagonal bands represent the region at $1\sigma$ (gray) and $3\sigma$ (light-gray) statistical confidence levels obtained by the linear fit of the data.  Open downward triangles locate the BA-type magnetic stars not detected by the VLA observations reported in this paper (below the $3\sigma$ detection threshold); the downward arrowhead corresponds to the luminosity estimated using the RMS measured on the radio maps ($1\sigma$ level).
}
\label{fig:cbo_vs_radio}
\end{figure}

The distributions of stellar parameters of all the BA-type magnetic stars reported in this paper, as well as the other stars discovered to be radio loud and with known parameters, which have already been analyzed in Paper\, I, in \citet{Shultz2022}, and in \citet{Das_Driessen2025}, are displayed in Fig.\,\ref{fig:isto_all_param_sample}. Note that most of the undetected stars have low temperatures. 
For those that are radio loud, the four parameters $B_{\text p}$, $R_{\ast}$, $P_{\text {rot}}$, and $M_{\ast}$ have been used to calculate the power produced by the CBOs, using the Eq.\,\ref{eq:lrad_lcbo}.
Their spectral radio luminosities (erg\,s$^{-1}$\,Hz$^{-1}$) are reported in Fig.\,\ref{fig:cbo_vs_radio} as a function of the correspondent $L_{\text{CBO}}$ (erg\,s$^{-1}$) values. For the sample analyzed in this paper, radio measurements are listed in Table~\ref{tab:fluxes}. A total of 79 radio loud BA-type magnetic stars are displayed in the $L_{\nu, {\mathrm {rad}}}-L_{\mathrm{CBO}}$ diagram. The median value of the CBO's power of the detected BA-type magnetic stars reported in the figure is $\langle L_{\mathrm{CBO}} \rangle \approx 10^{35.45}$~(erg\,s$^{-1}$). 
The high positive correlation between $L_{\nu, {\mathrm {rad}}}$ and $L_{\mathrm{CBO}}$ is confirmed by the values of the Pearson's linear correlation coefficient ($\approx 0.86$) and Spearman's correlation coefficient ($\approx 0.84$). The proportionality coefficient between the two luminosities obtained by the simple linear fit of the data is $10^{-19.2\pm0.4}$\,(Hz$^{-1}$); therefore, the scaling relationship between the two parameters reported in Fig.\,\ref{fig:cbo_vs_radio} is: 
\begin{equation}
\label{eq:empiric_scal_rel}
L_{\nu, {\mathrm {rad}}}=10^{-19.2\pm0.4} L_{\mathrm{CBO}}. 
\end{equation}
The grey band displayed in Fig.\,\ref{fig:cbo_vs_radio} represents the $1 \sigma$ uncertainty region, whereas the light-grey wider band refers to the $3 \sigma$ dispersion data. Based on the scaling relationship, the average  $\langle  L_{\mathrm{CBO}} \rangle$ corresponds to the spectral radio luminosity $\langle  L_{\nu, {\mathrm {rad}}}\rangle \approx 10^{16.25}$~(erg\,s$^{-1}$\,Hz$^{-1}$).

The figure also reports the non-detected stars analyzed in this paper (displayed using the open symbol). As it is evident, the upper limits of the $L_{\nu, {\mathrm {rad}}}$ (estimated by the $3\sigma$ level measured on the radio maps) of the undetected sources (open symbols) {fall mostly within} the light-grey region; therefore, we cannot certify such stars as definitively radio quiet. A necessary condition for magnetic BA-type stars to emit non-thermal radio is the presence of ionized material around them. As discussed in Sec.\,\ref{sec:scal_rel}, the stellar wind is the most obvious source of the magnetospheric plasma of BA-type stars. Based on the correlation between the mass loss rate of the radiatively driven stellar wind and the stellar temperature \citep{Krticka2014}, the mass loss rate of the wind is expected to be stronger if the stellar temperature is higher. This suggests a possible dependence of the radio emission level on the amount of magnetospheric plasma deposited by the wind.

\subsection{The $\log g$--$T_{\mathrm{eff}}$ diagram}
\label{sec:res_logg_logt}

\begin{figure}[]
\centering
\includegraphics[width=0.89\columnwidth]{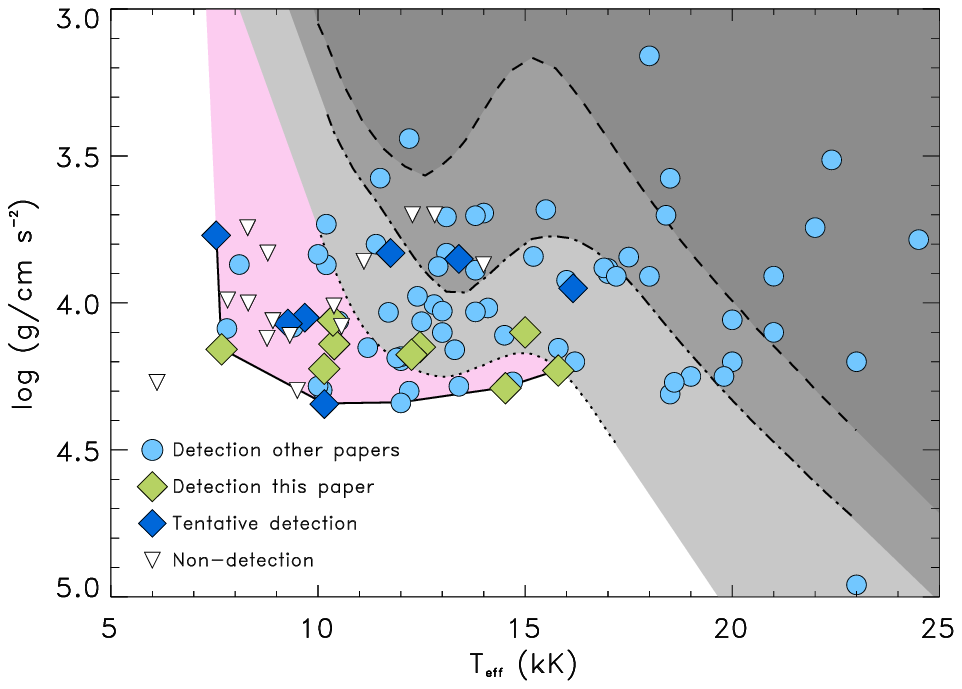}
\vspace{-2. mm}
\caption{Diagram $\log g$--$T_{\mathrm{eff}}$. Dashed and dot-dashed black lines are taken from \citet{Babel1996}, the dotted line is taken from  \citet{Hunger_Groote1999}. These lines locate the zone where stellar winds are theoretically expected. Dark grey zone: homogeneous wind; grey zone: multicomponent wind; light grey zone: multicomponent wind with a lower fraction of hydrogen coupled to metals.
These zones are obtained by extrapolating the original curves to encompass a range of parameters large enough to include all the BA-type magnetic stars reported in this paper, as well as other stars already discovered to be radio-loud. The pink zone, delimited by the black solid line (and its extrapolations), identifies a region on the $\log g$--$T_{\mathrm{eff}}$ parameter's space that is outside the wind zones but where several radio-loud stars still fall. The fact that there are detected stars outside the wind zone perhaps means that the wind zones do not have robust boundaries. On the other hand, the presence of undetected stars in the wind regions requires additional highly sensitive radio observations to see whether these stars are indeed radio quiet or have only very faint radio emissions.
}
\label{fig:logg_teff}
\end{figure}

The radiatively driven stellar wind of the B-type stars has been widely studied from a theoretical point of view. In particular, the stellar gravity effect in opposing to the radiative acceleration provided by the radiation field of the stellar photosphere has been parametrized by locating the {analyzed} star inside the diagram where the two parameters: gravity ($\log g$) and temperature ($T_{\text{eff}}$) are reported \citep{Babel1996,Hunger_Groote1999}. From a qualitative point of view, hot stars with low gravity at the surface have the most favorable conditions to originate homogeneous wind (dark-grey zone in Fig.\,\ref{fig:logg_teff}). At the lower temperature and higher gravity, multi-component metallic winds can still occur (grey and light-grey zones in Fig.\,\ref{fig:logg_teff}).

To better understand the general conditions governing the radio emission of the early-type magnetic stars, we look for a possible dependence of the radio-loud nature of the BA-type magnetic stars on their ability to lose ionized material by stellar wind. To do this, we locate all the stars reported in Fig.\,\ref{fig:cbo_vs_radio} inside the $\log g$ -- $T_{\mathrm{eff}}$ diagram shown in Fig.\,\ref{fig:logg_teff}. Looking at the figure, the larger number of BA-type magnetic stars lie within the theoretically predicted wind zones. But a large number of magnetic stars that are radio loud fall outside the boundary of the diagram wind zone (pink shaded area of Fig.~\ref{fig:logg_teff}). Their nature as a radio source definitely certifies them as stars surrounded by a magnetosphere; therefore, due to the limitation of a definitive understanding of the wind-driving conditions for late B and A stars, we evidenced that the wind-zone boundary gives only rough constraints regarding the effective presence of plasma surrounding the stars with temperatures and gravities that locate them outside the wind-zones of the $\log g$ -- $T_{\mathrm{eff}}$ diagram. Further, most of the undetected stars fall in the pink transition region. This is reasonable to expect because the non-detected stars are mostly distributed at the lower side of the temperature and mass distributions (see first and fourth panels of Fig.\,\ref{fig:isto_all_param_sample}). 

\section{Discussion}
\label{sec:discussion}

Beyond the BA-type magnetic stars, other classes of stars exist that are surrounded by {large-scale, well-ordered magnetospheres \citep{Donati2006,Reiners_Basri2007},} which are sites of plasma processes responsible for incoherent non-thermal radio emission and coherent highly-polarized radio emission of auroral origin. These are the Ultra Cool Dwarfs (UCDs), which are stellar and sub-stellar objects with spectral types later than M7 \citep{Pineda2017} {with suppressed coronal magnetic activity \citep{Mclean2012,Williams2014}.}

As evidenced in Paper\,I, when the incoherent non-thermal radio emission component from the UCDs is detected \citep{Metodieva2017}, these stars seem to have a behavior of their incoherent radio emission which follows the scaling relationship that applies to early-type magnetic stars. 
{Recently, a comparison of the plasma effects underlying the radio behavior of these extreme classes of stars \citep{Das_Owocki2025} has been performed, highlighting differences likely related to the low density of the ambient plasma. Indeed, unlike early-type stars, UCDs cannot produce plasma on their own due to their extremely weak or nonexistent winds.}



Despite the high number of UCDs located in the Sun's neighbourhood, a low number of them are detected as a source of incoherent radio emission \citep{Kao2024}. The low detection rate could be related to the external origin of the plasma source filling their magnetospheres, with consequent possible wide differences in magnetospheric plasma density between UCDs having similar magnetic field strength and rotation speed. 

From a qualitative point of view, the plasma density could affect the level of incoherent radio emission of the stars with well-ordered magnetospheres. Exploiting the BA-magnetic stars as an astrophysical plasma laboratory, we quantitatively investigate how the thermal plasma density of the equatorial {plasma disk} may affect the location where centrifugal breakouts occur and the subsequent possible influence on the gyro-synchrotron radio emission produced by relativistic electrons accelerated in the current sheets originating from the CBOs.

\subsection{A simplified approach to locating the CBO's site}
\label{sec:cbo_location}

The scenario that qualitatively describes the mechanism supporting the radio emission from the magnetospheres of the BA-type stars is pictured in Fig.~\ref{fig:scenario_model}. The centrifugal breakouts paradigm explains the non-thermal electrons present within their magnetospheres. The detailed study of the magnetospheres of early-type stars is not a simple matter; their modeling has been approached by solving the equation of motion governing the dynamics of the charged particles of the stellar wind channeled by the presence of the magnetic field \citep{Townsend_Owocki2005,Townsend2007,Owocki_Cranmer2018,Owocki2020,Berry2022}, also involving MHD simulations, 2D \citep{ud-doula2002,ud-doula2006,ud-doula2008,ud-Doula2014} and 3D \citep{ud-Doula2013,Daley-Yates2019,ud-Doula2023}.

The MHD detailed study of the CM of the early-type magnetic star is the optimal way for studying individual objects. But to investigate the average behavior of a large number of such kind of stars, an MHD approach requires a huge computational effort, which we avoid by following a simplified approach to roughly estimate the radial distance ($R_{\text{CBO}}$) at which the CBOs occur, that is, reasonably, the site where the energetic electrons are injected within the stellar magnetosphere. We calculated the energy density of an elementary volume of plasma located in the equatorial plane, assumed with zero thickness, taking into account separately the rotational ($u_{\Omega}$), gravitational ($u_{\text G}$), and thermal ($u_{\text T}$) terms, and then comparing with the magnetic energy density ($u_{\text B}$). For the calculation, we assumed the simplified hypothesis of an aligned dipole field (i.e., magnetic and rotation axes are coincident). Therefore, we neglect any longitudinal effects on the plasma density law at the magnetic equatorial plane. 

The centrifugal energy density of the material accumulated in the equatorial corotating disk is calculated by the equation:
\begin{displaymath}
u_{\Omega}(r)=\frac{1}{2} \boldsymbol {\rho(r)} \Bigg( \frac{2\pi r}{P_{\text{rot}}} \Bigg)^2,  
\end{displaymath}
the gravitational energy density is:
\begin{displaymath}
u_{\text G}(r)={ -G M_{\ast} \rho(r)\Bigg(  \frac{1}{r} \Bigg),}
\end{displaymath}
where the density decreases outward following a general radial density law: 
\begin{equation}
\label{eq:dens_law}
\boldsymbol {\rho(r)}=m_{\text p} n_0 \Bigg(  \frac{R_{\ast}}{r} \Bigg)^{q},  
\end{equation}
with the plasma considered pure, fully ionized Hydrogen (with $m_{\text p}$ proton mass; $n_0$ electron density number at the surface). 

The thermal energy density $u_{\text T}$ is simply the plasma pressure. Assuming the plasma of the disk is in nearly isobaric condition \citep{ud-Doula2014}, following the MCWS model, the radial decrease of the plasma density is compensated by the increase in plasma temperature \citep{Babel1997}, hence, the plasma pressure along the disk can be rougly quantified by the thermal energy density $u_{\text T} = P =n_0 k_{\text B} T_{\text{eff}}$ ($k_{\text B}$ Boltzman's constant) estimated at the stellar surface.

\begin{figure}[]
\centering
\includegraphics[scale=0.64]{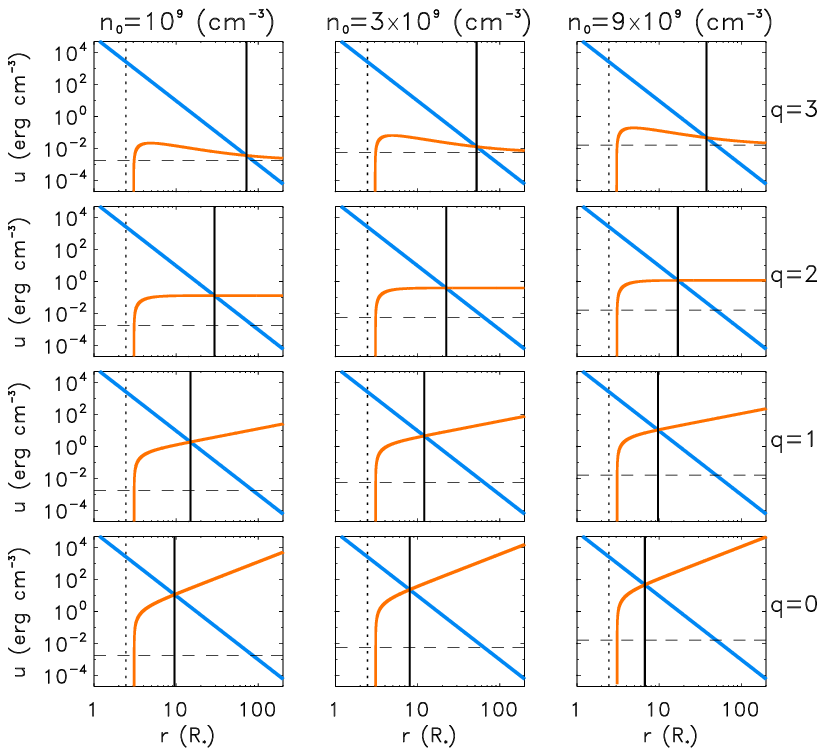}
\vspace{-5.5 mm}
\caption{Blue solid line shows the magnetic energy density radial distribution. The red solid lines illustrate the radial dependence of the combined effects of centrifugal, gravitational, and thermal energy densities. The vertical solid line locates the radial distance where the blue and red lines intersect. Stellar parameters refer to the reference star, which has stellar parameters where the distributions peak ($T_{\text{eff}}=13$\,kK; $R_{\ast}=2.7$\,R$_{\odot}$; $M_{\ast}=3.2$\,M$_{\odot}$; $B_{\text p}=3.1$\,kG; $P_{\text{rot}}=1.1$\,d). The thermal plasma pressure, evaluated under isobaric conditions, is marked by the horizontal dashed line. The vertical dotted line marks the position of the Kepler radius. Different cases of plasma density and the surface and exponent of the corresponding radial density law have been considered.
}
\label{fig:rcm_simul_stars}
\end{figure}

\begin{figure*}[]
\centering
\includegraphics[scale=0.81]{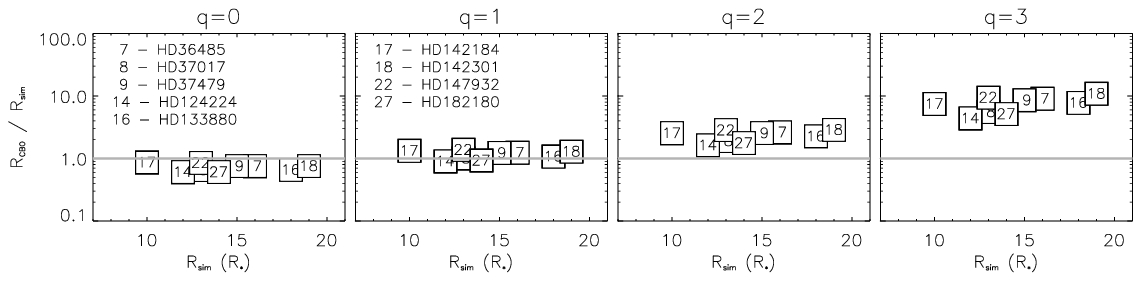}
\vspace{-2 mm}
\caption{Comparison between the location where the non-thermal electrons are injected within the magnetosphere of BA-type magnetic stars as calculated in this paper ($R_{\text{CBO}}$) and estimated in Paper\,I to reproduce the observed spectra of several BA-type magnetic stars well studied at the radio regime  ($R_{\text{sim}}$). Different cases of exponents ($q$) of the radial density law of the equatorial thermal plasma have been used to calculate $R_{\text{CBO}}$ (Eq.\,\ref{eq:r_cbo}). The grey line in each panel sets the condition $R_{\text{CBO}}=R_{\text{sim}}$.}
\label{fig:rapp_rr_simul_stars}
\end{figure*}

For the calculation of the magnetic energy density along the equatorial radial distance, we exploited the results retrieved by the detailed modeling of early-type magnetospheres. The magnetic field lines stretching, produced by the centrifugal force acting on the equatorial plasma co-rotating with the magnetic field \citep{ud-doula2002,ud-doula2006,ud-doula2008,ud-Doula2014}, can reasonably be expected to induce changes in the radial dependence of the magnetic field strength of a simple dipole. In general, at the magnetic equator, the radial dependence of the magnetic field strength is defined by the following relation:
\begin{equation}
\label{eq:campo_mag_monopolo}
B(r)=B_{\ast}\Bigg(  \frac{R_{\ast}}{r} \Bigg)^{p+1}, 
\end{equation}
where $B_{\ast}=B_{\text p}/2$ and $p=1$, which well describes the monopole topology of the stretched equatorial magnetic field of stars with CMs \citep{Owocki2022}. Using the above relation, the magnetic energy density at the magnetic equatorial plane depends on the radial distance, in the cgs system, as follows:
\begin{displaymath}
u_{\text B}(r)=\frac{B(r)^2}{8\pi} = \frac{B_{\text p} ^2}{32\pi} \Bigg(  \frac{R_{\ast}}{r} \Bigg)^{4}. 
\end{displaymath}
The radial distance of the site where CBOs occur can be roughly quantified by simply equating the magnetic energy density to the centrifugal plus thermal energy density, reduced by the gravitational attracting effects. The distance of the equatorial magnetospheric region where the following equality is {satisfied}:
\begin{equation}
\label{eq:r_cbo}
u_{\text B}= u_{\Omega}+u_{\text T} {+u_{\text G}}
\end{equation}
roughly locates the breakout region. 

Several values of the exponent ($q$) of the radial density law (Eq.\,\ref{eq:dens_law}) have been analyzed, varying from a homogeneous matter distribution in the equatorial {plasma disk}, flat density law ($q=0$), up to the steepest decrease: $q=3$. Once the value of $q$ is assigned, we numerically estimated the distance where the equality between the two members of Eq.\,\ref{eq:r_cbo} is verified. We analyzed three different values of $n_0$, respectively: $n_0=10^9$\,cm$^{-3}$; $n_0=3\times10^9$\,cm$^{-3}$; $n_0=9\times10^9$\,cm$^{-3}$. The radial profiles of the two sides of Eq.\,\ref{eq:r_cbo} are displayed in Fig.\,\ref{fig:rcm_simul_stars}. For the calculation, we used stellar parameters that are considered typical for the BA-type radio-loud magnetic stars. The adopted parameters are those where the distributions shown in Fig.\,\ref{fig:isto_all_param_sample} peak. In particular: $T_{\text{eff}}=13$\,kK; $R_{\ast}=2.7$\,R$_{\odot}$; $M_{\ast}=3.2$\,M$_{\odot}$; $B_{\text p}=3.1$\,kG; $P_{\text{rot}}=1.1$\,d. 

Once the polar field strength and the stellar radius are assigned, the magnetic energy density radial profile is {uniquely} defined, that is, the first {term} of Eq.\,\ref{eq:r_cbo}. Whereas the terms of the second member of Eq.\,\ref{eq:r_cbo} depend on $n_0$ and $q$. The solid vertical line pictured in each panel of Fig.\,\ref{fig:rcm_simul_stars} locates the distance at which the magnetic energy density (solid blue line) equals the second member of Eq.\,\ref{eq:r_cbo} (solid red line).

Looking at Fig.\,\ref{fig:rcm_simul_stars}, it is evident that for all the combinations of adopted parameters, the regions where the CBOs occur are located at distances larger than $R_{\text K}$ (marked by the vertical dotted line in each panel of the figure), which for the adopted stellar parameters is about 2.4 stellar radii large. At distances lower than $R_{\text K}$, the gravitational terms dominate over the centrifugal and thermal ones, making negative the second member of Eq.\,\ref{eq:r_cbo}.

The centrifugal effect acting on the {plasma disk} depends on the density at the surface ($n_0$) and the exponent of the radial density law ($q$). In {plasma disks} that are less dense and that have a steeper radial decrease (panel at the top left corner of Fig.\,\ref{fig:rcm_simul_stars}) the CBOs region is located far from the star, whereas this is closer in cases of higher density disk and with a flatter radial decrease (panel at the bottom right corner of the figure). We also noticed that the plasma thermal pressure is not negligible only in cases of the steepest radial density law that we have considered (top panels of Fig.\,\ref{fig:rcm_simul_stars}).

\begin{table*}
\caption[ ]{Input parameters ($P_{\text{rot}}$, $B_{\text{p}}$, and $n_0$) and solution of the reletivistic electrons column density ($n_{\text r}\times l_{\text{}}$). The fixed parameters of the reference star are: $R_{\ast}=2.7$\,R$_{\odot}$, $M_{\ast}=3.2$\,M$_{\odot}$. The exponent of the radial density law of the {plasma disk} is $q=1$.}
\vspace{-5. mm}
\label{tab:param_simul}
\footnotesize
\begin{center}
\begin{tabular}{c c c r cc r cc r cc}                         
\hline                           
\hline    
\multicolumn{12}{l}{$B_{\text{p}}=3100$\,(G)}       \\
\hline    

~ &~ &~ &~ &\multicolumn{2}{c}{$n_0=10^9$\,(cm$^{-3}$)} &~ &\multicolumn{2}{c}{$n_0=3\times 10^9$\,(cm$^{-3}$)} &~ &\multicolumn{2}{c}{$n_0=9\times 10^9$\,(cm$^{-3}$)} \\
\cline{5-6}
\cline{8-9}
\cline{11-12}
$P_{\text{rot}}$ &$L_{\text{CBO}}$ &$L_{\nu,\text{rad}}$       &~ &$R_{\text{CBO}}$ &$n_{\text r} \times l_{\text{}}$ &~ &$R_{\text{CBO}}$ &$n_{\text r} \times l_{\text{}}$ &~ &$R_{\text{CBO}}$ &$n_{\text r} \times l_{\text{}}$   \\
(d)              &(erg\,s$^{-1}$)  &(erg\,s$^{-1}$\,Hz$^{-1}$) &~ &(R$_{\ast}$)     &(cm$^{-2}$)            &~ &(R$_{\ast}$)     &(cm$^{-2}$)            &~ &(R$_{\ast}$)     &(cm$^{-2}$)              \\

\cline{5-6}
\cline{8-9}
\cline{11-12}

\vspace{-2 mm}                                                                    
 &&&&&&&&&&& \\

0.11       &$10^{37.44}$ &$\approx 10^{18.2}$ &~ &$\approx 6~~~$    &$2.04\times 10^{16}$ &~ &$\approx 5~~~$    &$1.76\times 10^{16}$ &~ &$\approx 4~~~$     &$1.41\times 10^{16}$   \\
0.35       &$10^{36.44}$ &$\approx 10^{17.2}$ &~ &$\approx 9.5$     &$2.94\times 10^{15}$ &~ &$\approx 7.5$     &$2.25\times 10^{15}$ &~ &$\approx 6~~~$     &$2.11\times 10^{15}$   \\
1.1~~      &$10^{35.44}$ &$\approx 10^{16.2}$ &~ &$\approx 15~~~~~$ &$4.27\times 10^{14}$ &~ &$\approx 12~~~~~$ &$4.40\times 10^{14}$ &~ &$\approx 9.5$      &$3.02\times 10^{14}$   \\
3.5~~      &$10^{34.44}$ &$\approx 10^{15.2}$ &~ &$\approx 24~~~~~$ &$6.26\times 10^{13}$ &~ &$\approx 19~~~~~$ &$5.64\times 10^{13}$ &~ &$\approx 15.5~~$   &$4.23\times 10^{13}$   \\
11~~~~~~~  &$10^{33.44}$ &$\approx 10^{14.2}$ &~ &$\approx 37.5~~$  &$1.04\times 10^{13}$ &~ &$\approx 30~~~~~$ &$8.81\times 10^{12}$ &~ &$\approx 24~~~~~$  &$6.50\times 10^{12}$   \\


\hline   

\multicolumn{12}{l}{$P_{\text{rot}}=1.1$\,(d)}       \\
\hline

~ &~ &~ &~ &\multicolumn{2}{c}{$n_0=10^9$\,(cm$^{-3}$)} &~ &\multicolumn{2}{c}{$n_0=3\times 10^9$\,(cm$^{-3}$)} &~ &\multicolumn{2}{c}{$n_0=9\times 10^9$\,(cm$^{-3}$)} \\
\cline{5-6}
\cline{8-9}
\cline{11-12}
$B_{\text{p}}$   &$L_{\text{CBO}}$ &$L_{\nu,\text{rad}}$       &~ &$R_{\text{CBO}}$ &$n_{\text r} \times l_{\text{}}$ &~ &$R_{\text{CBO}}$ &$n_{\text r} \times l_{\text{}}$ &~ &$R_{\text{CBO}}$ &$n_{\text r} \times l_{\text{}}$   \\
(G)              &(erg\,s$^{-1}$)  &(erg\,s$^{-1}$\,Hz$^{-1}$) &~ &(R$_{\ast}$)     &(cm$^{-2}$)            &~ &(R$_{\ast}$)     &(cm$^{-2}$)            &~ &(R$_{\ast}$)     &(cm$^{-2}$)              \\

\cline{5-6}
\cline{8-9}
\cline{11-12}

\vspace{-2 mm}                                                                    
 &&&&&&&&&&& \\

31000     &$10^{37.44}$ &$\approx 10^{18.2}$ &~ &$\approx 37.5~~$  &$1.56\times 10^{16}$ &~ &$\approx 30~~~~~$   &$1.41\times 10^{16}$ &~ &$\approx 24~~~~~$   &$1.20\times 10^{16}$  \\
10000     &$10^{36.44}$ &$\approx 10^{17.2}$ &~ &$\approx 24~~~~~$ &$2.35\times 10^{15}$ &~ &$\approx 19~~~~~$   &$2.14\times 10^{15}$ &~ &$\approx 15.5~~$    &$2.11\times 10^{15}$  \\
~~3100    &\multicolumn{3}{c}{Parameters already listed}  &-- &-- & &-- &-- & &-- &--  \\
~~1000    &$10^{34.44}$ &$\approx 10^{15.2}$ &~ &$\approx 9.5$     &$7.83\times 10^{13}$ &~ &$\approx 7.5$       &$4.70\times 10^{13}$ &~ &$\approx 6~~~$      &$2.82\times 10^{13}$  \\
~~~~310   &$10^{33.44}$ &$\approx 10^{14.2}$ &~ &$\approx 6~~~$    &$9.40\times 10^{12}$ &~ &$\approx 5~~~$      &$4.70\times 10^{12}$ &~ &$\approx 4~~~$      &$2.82\times 10^{12}$  \\


\hline    



\end{tabular}

\end{center}

\end{table*}

To search for an indirect probe of what radial density law can better describe the true radial dependence of the ionized material accumulated within the equatorial {plasma disk}, we exploited the results derived by the simulations of a sub-sample of stars with well-studied wide-band radio spectra, performed in Paper\,I. In Fig.\,\ref{fig:rapp_rr_simul_stars} we show the ratio between the theoretically expected location where the relativistic electrons are injected ($R_{\text{CBO}}$) and the corresponding injection distance ($R_{\text{sim}}$) required to reproduce the observed radio spectra of the stars simulated in Paper\,I, as a function of $R_{\text{sim}}$ (these radii are listed in Table\,2 of Paper\,I). The stars are labeled by using the same identification number used in Table\,1 of Paper\,I.  In Paper\,I it was assumed the same value of the thermal electron density at the stellar surface, that is $n_0=3\times10^9$\, cm$^{-3}$, for each source. Therefore, the same density has been adopted to numerically estimate the $R_{\text{CBO}}$ of these stars by using Eq.\,\ref{eq:r_cbo}. Looking at Fig.\,\ref{fig:rapp_rr_simul_stars}, a radial linear decrease ($q=1$) of the density of the equatorial {plasma disk} seems to be in better agreement with the location where the non-thermal electrons are injected, as required to reproduce the observed spectra of the sub-sample of stars analyzed in Paper\,I. Note that, 
{the radial density law reported here, which was empirically constrained, provides a rough description of the plasma disk at large distances (namely $r \gg R_{\text K}$), where the disk scale height asymptotically approaches a constant value \citep{Townsend_Owocki2005}. The radial density law reported above cannot be applied approaching the Kepler radius, where the surface density of the plasma disk evidenced a steeper radial decline \citep{Townsend_Owocki2005,Owocki2020,Berry2022,ud-Doula2023}.}




From the above analysis, we excluded HD\,147933 ($\rho$\,Oph\,A), which was previously included in Paper\,I. This magnetic B-type star has been recently discovered as a binary system composed of two B-type stars \citep{Shultz2025}. The magnetized one, and also radio loud, is the secondary. The stellar parameters of this star have been changed concerning the values used in Paper\,I, mainly the stellar radius and magnetic field strength. The revised parameters make the results of the simulation of the radio spectrum of HD\,147933 performed in Paper\,I unreliable; therefore, we decided to exclude this star from the above comparison between $R_{\text{CBO}}$ and $R_{\text{sim}}$.

\subsection{3D model radio spectrum calculation}
\label{sec:spectrum_calculation}

Using the method to locate the CBO described in Sec.\,\ref{sec:cbo_location}, we calculate the magnetospheric gyro-synchrotron radio emission exploring the parameter space on which it depends. The 3D model of gyro-synchrotron radio emission from a dipole-shaped magnetosphere \citep{Trigilio_etal2004,Leto2006} (which was used in Paper\,I to simulate the measured radio spectra of some BA-type magnetic stars well studied at the radio wavelengths) is used to calculate synthetic radio spectra of a star with average parameters of the whole sample of stars reported in Fig.\,\ref{fig:cbo_vs_radio}.

{We performed this exercise using  the stellar parameters where the distributions shown in Fig.\,\ref{fig:isto_all_param_sample} peak.}
The reference star has $L_{\text{CBO}} \approx 10^{35.44}$\,erg\,s$^{-1}$, calculated using the Eq.\,\ref{eq:lrad_lcbo}, which coincides almost perfectly with the median value of all the stars reported in Fig.\,\ref{fig:cbo_vs_radio} (see Sec.\,\ref{sec:obs_results}). The expected spectral radio luminosity (Eq.\,\ref{eq:empiric_scal_rel}) is $L_{\nu,\text{rad}} \approx 10^{16.2}$\,erg\,s$^{-1}$\,Hz$^{-1}$. We calculated the corresponding radio spectra as a function of the thermal electron density of the {plasma disk} at the stellar surface (same values of $n_0$ used in Sec.\,\ref{sec:cbo_location}). 

To account for the rotational modulation effect of the radio emission from the corotating magnetosphere, we adopted a dipole tilted by $45^{\circ}$ to the rotation axis, assumed perpendicular to the line of sight. Two different magnetospheric orientations have been chosen for calculating the wide-band synthetic radio spectra (explored spectral range: 0.1--200\,GHz), respectively: the equator on view, corresponding to a null of the effective magnetic field, and the orientation that makes the pole better visible, corresponding to an extremum of the effective magnetic field curve. The adopted oblique rotator model geometry produces a symmetric magnetic field curve; therefore, the viewing of the south or north magnetic pole does not influence the total intensity (Stokes\,I) of the gyro-synchrotron radio emission. 

We considered the average spectrum between the two, calculated for different magnetospheric orientations. The simulated radio spectra exhibited a steep behavior at both lower and higher radio frequencies (see middle panels of Figs.\,\ref{fig:spe_maps_simul_Bfixed} or \ref{fig:spe_maps_simul_Pfixed}). To reproduce the required radio luminosity, we selected the spectral region where the average spectrum is approximately flat, varying the unique free parameter of the model: the relativistic electron column density. This is the product of the relativistic electrons' number density ($n_{\text r}$\, cm$^{-3}$) and the equatorial length of the region where the centrifugal breakouts occur ($l_{\text{}}$\,cm). The value of $n_{\text r}\times l_{\text{}}$ that reproduces the radio luminosity level is listed in Table\,\ref{tab:param_simul}. 

\begin{figure}[]
\centering
\includegraphics[width=1.35\columnwidth]{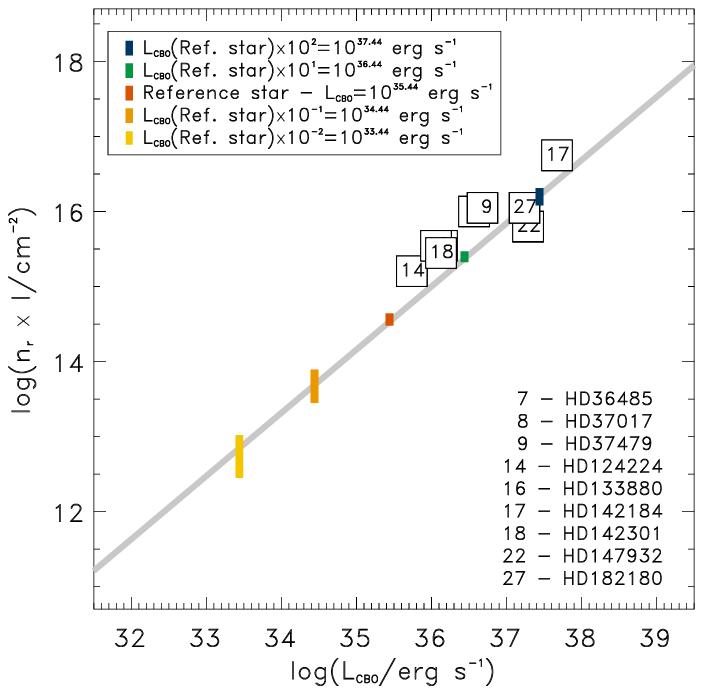}
\vspace{-5 mm}
\caption{Column density of the relativistic electrons reproducing the radio emission level expected from a fixed value of the power released by the CBOs. The calculations were performed considering different combinations of stellar parameters capable of providing the same value of $L_{\text{CBO}}$; this explains the different lengths of the colored bars, which represent the derived values of $n_{\text r}\times l$ (Table\,\ref{tab:param_simul}). The grey line is the best fit of the two parameters. Open square symbol: column densities that reproduce the observed radio spectra of the stars studied in Paper\,I.
}
\label{fig:nrl_per_l}
\end{figure}

\begin{figure*}[]
\centering
\includegraphics[width=1.8\columnwidth]{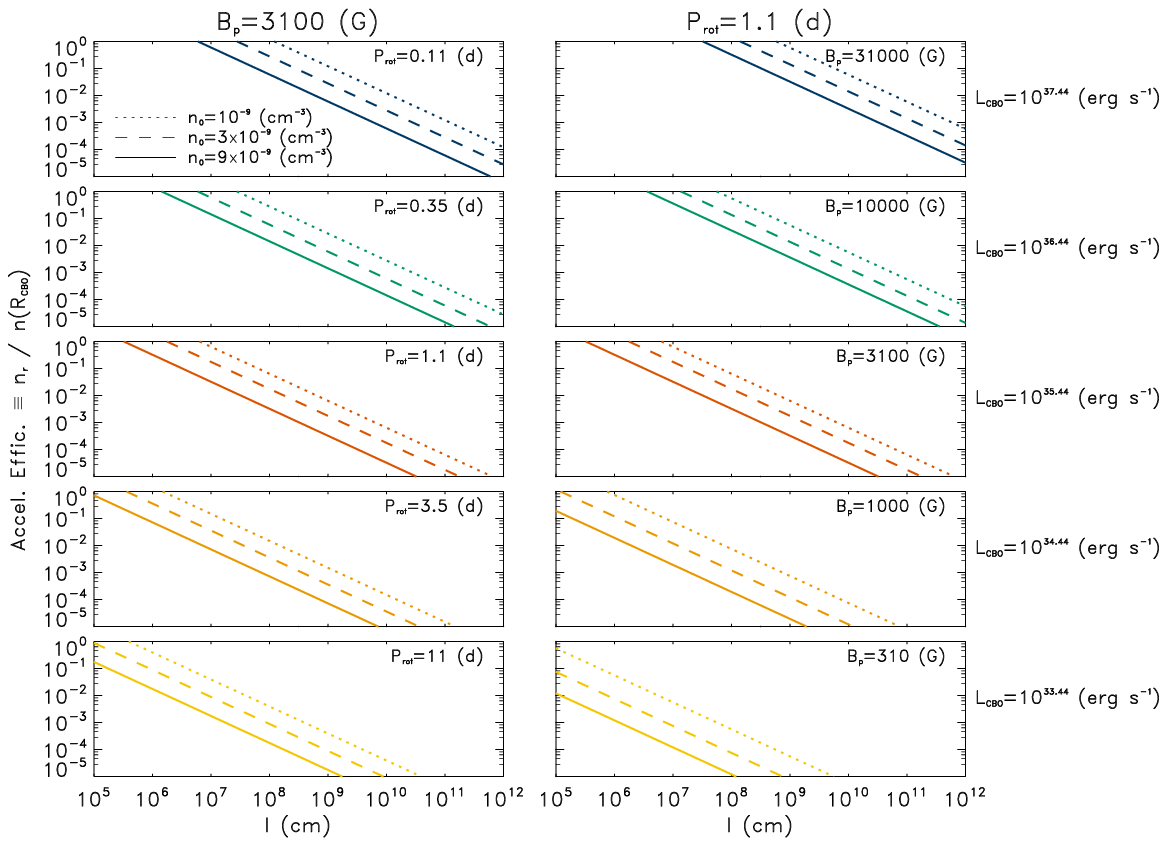}
\vspace{-1. mm}
\caption{Required efficiency of the non-thermal electron production to reproduce the radio emission level expected from the $L_{\nu,{\text{rad}}}$ vs  $L_{\text{CBO}}$ relationship, as a function of the length of the equatorial current sheet that originates as a consequence of CBOs. Each panel reports the results obtained using three values of the thermal electron density of the equatorial {plasma disk} at the stellar surface for a fixed combination of stellar rotation period and polar magnetic field strength. 
}
\label{fig:effic_cbo}
\end{figure*}

To explore the behaviour of the BA-type magnetic stars distributed over the range of CBO powers covered by the stars reported in Fig.\,\ref{fig:cbo_vs_radio}, we studied the effects on the expected radio emission separately varying $P_{\text{rot}}$ and $B_{\text p}$ within ranges wide enough to cover a range of 2 magnitude higher and lower of the expected $L_{\text{CBO}}$ respect the median value. These two model parameters, together with the other explored parameter: the plasma density of the {plasma disk} at the stellar surface, are listed in Table\,\ref{tab:param_simul} too. The  stellar mass and radius were left fixed ($R_{\ast}=2.7$\,R$_{\odot}$;  $M_{\ast}=3.2$\,M$_{\odot}$). Once the combination of the input parameters is assigned, the calculations of the radio spectra are performed following the same modeling approach described above. The synthetic radio spectra calculated using the parameters listed in Table\,\ref{tab:param_simul}, are shown in Figs.\,\ref{fig:spe_maps_simul_Bfixed} and \ref{fig:spe_maps_simul_Pfixed}. The individual spectra behavior is discussed in Appendix\,\ref{appendix:ref_star_spectra}. 

The distribution of the distances of the whole sample of BA magnetic stars pictured in the first panel of Fig.\,\ref{fig:isto_all_param_sample} peaks at 125\,pc. In Table~\ref{tab:param_simul_appendix} we listed the flux levels measurable at Earth corresponding to the radio luminosities listed in Table\,\ref{tab:param_simul} assuming the reference star 125\,pc away. In Table~\ref{tab:param_simul_appendix}, the magnetic field strengths at the distances where the CBO occurs are also listed.

\subsection{CBOs' efficiency for relativistic electrons production}
\label{sec:cbo_effic}

The values of $n_{\text r}\times l_{\text{}}$\,(cm$^{-2}$), derived by the simulations performed in Sec.\,\ref{sec:spectrum_calculation} and listed in Table\,\ref{tab:param_simul}, are reported as a function of the CBO power ($L_{\text{CBO}}$\,erg\,s$^{-1}$) in Fig.\,\ref{fig:nrl_per_l}. Although different combinations of stellar parameters producing the same expected $L_{\text{CBO}}$ require slightly different values of $n_{\text r}\times l$ to reproduce the expected radio luminosity, a clear correlation between column density and power of the CBOs exists. These two parameters are well fitted by a simple linear relation between their logarithms (pictured in the figure by the thick grey line), defined by the equation:
\begin{equation}
\label{eq:fit_nrl_lcbo}
\log \Bigg( \frac {n_{\text r}\times l} {\text{cm}^{-2}} \Bigg) = A + B\times \log \Bigg( \frac{L_{\text{CBO}}}{\text{erg\,s}^{-1}} \Bigg),
\end{equation}
where $A\approx -15.31$ and $B\approx 0.84$.

To compare the results of the simulations performed using the stellar parameters of reference stars with real cases, we exploit the results of the spectra simulations performed in Paper\,I. The relativistic electrons column density of the stars listed in Table\,2 of Paper\,I are reported in Fig.\,\ref{fig:nrl_per_l}. We noticed that all the stars {analyzed} in Paper\,I have $L_{\text{CBO}}$ higher than the median value ($L_{\text{CBO}}\approx 10^{35.4}$\,erg\,s$^{-1}$). This is not surprising because these stars are also those with higher radio emission levels. Therefore, they are easier to detect from Earth, allowing for obtaining well-sampled observed radio spectra. Furthermore, as demonstrated by \citet{Townsend_Owocki2005}, the tilted dipole magnetic configuration of a rigidly rotating magnetosphere influences the spatial distribution of the thermal plasma accumulation zone. Therefore, in cases of real magnetospheres, their individual ORM geometry can affect the true amount of non-thermal electron production and consequently the measurable radio emission level. In any case, although the ORM geometry of each star analyzed in Paper\,I is unique, their location in the diagram pictured in Fig.\,\ref{fig:nrl_per_l} is quite close to the best-fit line.

In cases of BA-type magnetic stars with centrifugal magnetospheres, the non-thermal electrons are likely to originate from the acceleration of thermal plasma provided by magnetic reconnection \citep{Li2015,Li2017,Zhang2021} expected to be occurring in the current sheet produced by the CBOs (see Fig.\,\ref{fig:scenario_model}). The amount of energetic electrons propagating within the magnetosphere surrounding a typical BA-type magnetic star is the key parameter in regulating its radio emission level. Based on the present study, we are then able to quantify the ratio between the number density of the relativistic electrons and the density of the local thermal plasma, that is the efficiency of the non-thermal electrons production, as a function of the length of the current sheet, which originates where the centrifugally supported thermal plasma of the magnetodisk breaks the magnetic field lines. This analysis has been performed for the three cases of the density of the equatorial {plasma disk} here analyzed. The results are shown in Fig.\,\ref{fig:effic_cbo}.

As a leading result of our study,  we notice that for the same efficiency, CBOs occurring in lower-density regions (dot coloured lines pictured in each panel of Fig.\,\ref{fig:effic_cbo}) always require a wider length of the particle acceleration site (i.e., the length $l_{\text{}}$ of the current sheet) to reproduce the radio emission level expected by the scaling relationship formulated by the Eq.\,\ref{eq:empiric_scal_rel}. Whereas, once the length $l$ is assigned, we noticed that a lower efficiency of the non-thermal electron production is systematically required when the density of the equatorial {plasma disk} is higher (solid coloured lines of Fig.\,\ref{fig:effic_cbo}). Conversely, the efficiency must increase as $n_0$ decreases to have the same radio emission level. This behavior seems unlikely.

It is counterintuitive to assume that the length of the acceleration region and the corresponding efficiency of the acceleration mechanism are inversely related to the density of the local thermal plasma. It is most logical to expect that the physical mechanism for the non-thermal electron production {works better if there are many free electrons in magnetic reconnection regions. Even assuming}
the worst condition, that is, the efficiency of the particle acceleration mechanism responsible for the non-thermal electrons production does not depend on the density of the local plasma where the magnetic reconnections occur, {the density of relativistic electrons is expected to increase as $n_0$ increases, with a consequent higher radio luminosity. The above consideration suggests that different stars having similar $L_{\text {CBO}}$, but with a plasma disk of different density, should have different levels of radio emission.} This would be an effect that introduces dispersion on the measured radio luminosities of stars with expected similar powers of the centrifugal breakouts.

\subsection{The non-detected BA-type magnetic stars}
\label{sec:non_detections}

In the previous section, we provided a recipe to roughly estimate the non-thermal electrons column density required to reproduce the radio luminosity predicted by Eq.\,\ref{eq:empiric_scal_rel} as a function of the power released by the centrifugal breakouts. Assuming that the BA-type magnetic stars reported in Fig.\,\ref{fig:cbo_vs_radio} have, on average, a similar efficiency of the acceleration mechanism for relativistic electron production, we look for common features that possibly differentiate the sub-sample of non-detected from the other stars that are radio loud. I.e., the non-detected stars seem to have distributions of distances, $B_{\text p}$, and $R_{\ast}$ similar to that of the radio-loud sample, but a longer $P_{\text{rot}}$ (Fig.\,\ref{fig:isto_all_param_sample}). As suggested by the dependence on
 centrifugal term in Eq.\,\ref{eq:r_cbo}, the rotation period plays a crucial role in locating the breakout region. 

\begin{figure}[]
\centering
\includegraphics[scale=0.52]{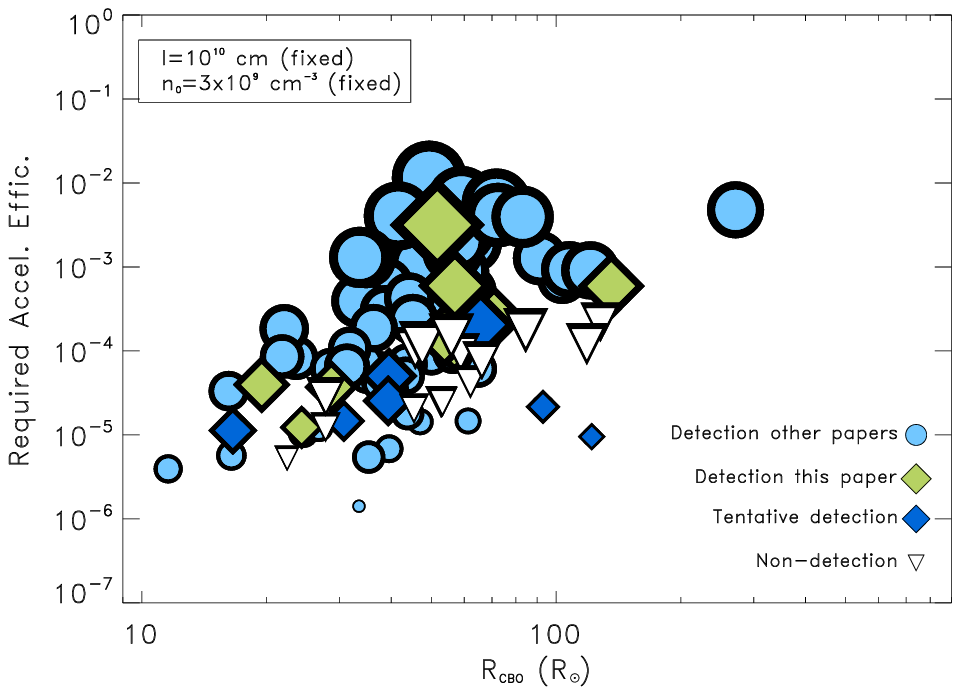}
\vspace{-2. mm}
\caption{Acceleration efficiency required to produce the measured radio emission level as a function of the equatorial distance where the energetic particles are accelerated. The size of the symbols is proportional to the $L_{\text{CBO}}$ of the individual star reported in Fig.\,\ref{fig:cbo_vs_radio}.
}
\label{fig:fig_ac_eff_all}
\end{figure}

We numerically estimated the location of the breakout region of all the stars shown in Fig.\,\ref{fig:cbo_vs_radio} by using Eq.\,\ref{eq:r_cbo}, that is the distance ($R_{\text{CBO}}$) at which the energetic particles responsible for the non-thermal radio emission are injected within the magnetosphere. Employing Eq.\,\ref{eq:dens_law}, we calculated the local density of the thermal electrons. Since this is a comparative analysis, we used the same value of the electron density of the equatorial disk at the stellar surface: $n_0=3\times10^{9}$\,cm$^{-3}$. Under the average behaviour summarized by Eq.\,\ref{eq:empiric_scal_rel}, the corresponding expected column density of the relativistic electrons was estimated for each star using Eq.\,\ref{eq:fit_nrl_lcbo}. 

For the Sun's flares, typical volumes of about $10^{30}$\,cm$^3$ of the coronal active regions where the flares originate have been reported \citep{Vlahos2004}. To estimate the required efficiency for the energetic particle acceleration within the magnetospheres of all the BA-type magnetic stars here analyzed, we always assume an effective acceleration length of about  $10^{10}$\,cm,  achieved by the study of the Sun's coronal flares. In Fig.\,\ref{fig:fig_ac_eff_all}, we reported the estimated efficiencies as a function of the $R_{\text{CBO}}$. 

The figure presents inconclusive results regarding the possibility of distinguishing between the non-detected stars and the others. On the other hand, a large number of undetected stars are colder than the average (see second panel of Fig.\,\ref{fig:isto_all_param_sample}) and are located in the transition region marked by the pink area in Fig.\,\ref{fig:logg_teff}. The analysis of BA-type magnetic stars in the context of their temperature and gravity suggests that the amount of magnetospheric thermal plasma could influence the level of incoherent radio emission. As a qualitative explanation for the observed superposition of detected and undetected stars with similar stellar parameters, the stars falling within this transition zone have a weak wind, resulting in low-density ionized material trapped within the stellar magnetosphere. This makes them faint radio sources, with an emission level of their incoherent emission component that should be difficult to detect. The stars located within the zone of the diagram $\log g$--$T_{\mathrm{eff}}$, where the conditions for the radiatively driven wind production are less favorable, will likely have a less dense {plasma disk}, with centrifugal breakouts occurring at larger distances, and consequently a lower (or undetectable) radio emission level. 

On the other hand, we cannot rule out that the non-detection could be the result of radio variability. In fact, the oblique rotator model geometry which characterizes such kind of magnetic stars gives rise to rotational variability of their gyro-synchrotron radio emission, which, in the case of single-epoch observations, could significantly affect the measurable radio emission level, making in the worst condition of measurements taken in coincidence with a minimum of the radio light curve, the stellar radio emission below the detection threshold.

\section{Summary and Conclusions}
\label{sec:conclusion}

In this paper, we report newly discovered BA-type magnetic stars that are radio loud. The well-known stellar parameters of the detected stars allowed us to calculate the expected power released by the CBOs. Locating these stars in the $L_{\nu,\text{rad}}$--$L_{\text{CBO}}$ diagram, we confirm that their radio behavior follows well the general scaling relationship of the stars with ordered magnetospheres.

The number of early-type magnetic stars detected at the radio regime is quickly increasing, and the new detections confirm that the scaling relationship between the power of the centrifugal breakouts and the spectral radio luminosity is a general law spanning five orders of magnitude (see Fig.\,\ref{fig:cbo_vs_radio}). On the other hand, we noticed that stars with similar $L_{\text{CBO}}$ show a not-negligible spread (about 1\,DEX)  of their measured spectral radio luminosities around the expected value.

Several reasons may affect the observable flux level. I.e., the selected observing frequency may fall outside the flat spectral region of the gyro-synchrotron radio spectrum. {The proper ORM geometry of the star has a role in the radio emission behavior. In fact, the magnetic dipole obliquity may affect the overall magnetospheric radio emission, as} discussed by \citet{Shultz2022} and \citet{Owocki2022}.


In this paper, we highlight another effect that is intimately related to the magnetospheric physical condition of the star, which in turn directly affects its corresponding radio emission level. This is the thermal plasma density of the magnetospheric regions where the energetic particles are accelerated. All the above-reported effects can contribute to increasing the dispersion around the mean behavior outlined by the scaling relationship illustrated in Fig.\,\ref{fig:cbo_vs_radio}. 

We hypothesized that in the CM magnetospheres of the BA-type magnetic stars, the centrifugal breakouts achieve plasma conditions similar to those occurring within the Solar corona during the impulsive flares, where magnetic reconnections and associated particle acceleration take place, but in the case of the CM magnetospheres, such plasma processes are steady.

New multi-frequency, rotationally averaged, highly sensitive radio observations using the next generation of ground-based radioastronomical observing facilities, mainly the next generation VLA (ngVLA) and the Square Kilometer Array (SKA), will be useful for a deep investigation of the plasma processes responsible for the production of non-thermal electrons triggered by the CBOs, and the related efficiency. The well-ordered magnetospheres surrounding early-type magnetic stars are very interesting scientific cases, with the added value of serving as laboratories for the study of astrophysical plasmas.

\begin{acknowledgements}{We thank the anonymous referee for his/her careful revision. AuD has received research support from NASA through Chandra Award number TM4-25001A issued by the Chandra X-ray Observatory 27 Center, which is operated by the Smithsonian Astrophysical Observatory for and on behalf of NASA under contract NAS8-03060.
JK was supported by grant GA \v{C}R 25-15910S. 
}
\end{acknowledgements}

%
%

\bibliographystyle{aa} 
\bibliography{biblio_paper}

\begin{appendix}




\onecolumn

\section{Stellar parameters of the selected sample}
\label{appendix:sample_param}

The fundamental stellar parameters of all the stars analyzed in this paper are listed in Table\,\ref{tab:star_param}.

\begin{table*}[h!]
\caption[ ]{Stellar parameters of the selected sample. The distances of almost all sources are the median of the photogeometric distance posterior given by \citet{Bailer-Jones2021}. The effective temperatures, stellar radii, and masses are mainly taken from the TESS Input Catalog v8.0 \citep{Stassun2019} and v8.2 \citep{Paegert2021}, and from \citet{Kervella2022}, whereas the polar magnetic field strength from \citet{Sikora2019a} and \citet{Shultz2022}.} 
\vspace{-4. mm}
\label{tab:star_param}
\footnotesize
\begin{center}
\begin{tabular}{@{}r@{~~}c@{~~} r@{~~~}                l@{~~~}                                     l@{~~~}              l@{~~~}              l@{~~~}                  l@{~~~}                             r@{}  }                         
\hline                                    
\hline                                    
           &$D$             &$T_{\mathrm{eff}}$~~~~~~  &~~~$\log g$                                &~~~~$R_{\ast}$      &~~~$M_{\ast}$       &$B_{\mathrm p}       $  &~~~~~~$P_{\mathrm{rot}}^{\dag}$    &$L_{\text{CBO}}$   \\
HD~~~~     &(pc)            &(kK)~~~~~                 &~~[cgs]                                    &~~~(R$_{\odot}$)    &~~(M$_{\odot}$)     &(kG)                    &~~~~~~(d)                          &(erg\,s$^{-1}$)   \\
\hline                                                            
\vspace{1 mm}                                
11502$^{\ast}$\, &53(1)$^{[1]}$  &10.4(0.2)~~~         &     $4.27_{-0.15}^{+0.1}    $$^{[4]}$     & 2.01(0.07)         &2.7(0.4)            &$3.0_{-0.75}^{+29.13}$  &~~n.a.                             & n.a.           \\    
\vspace{1 mm}                                                
12767~~         &106(1)          &13.4(0.2)~~~         &     $3.85(0.1)              $$^{[5]}$     & 3.5(0.2)           &4.1(0.2)            &$2.0_{-0.6}^{+0.5}   $  &~~1.892(n.a.)$^{[14]}$              & $1.11(0.69) \times 10^{35} $  \\    
\vspace{1 mm}                                                               
15089~~         &45.2(0.2)       & 8.8(0.1)~~~         &     $4.12_{-0.08}^{+0.07}   $$^{[4]}$     & 2.17(0.07)         &2.2(0.3)            &$1.85_{-0.16}^{+0.49}$  &~~1.74050(3)$^{[15]}$              & $1.78(0.36) \times 10^{34}$   \\    
\vspace{1 mm}                                                                 
21699~~         &177(4)          &12.5(0.2)~~~         &     4.15(n.a.)$^{[6]}$                    & 4.3(0.2)           &4.9(0.25)           &$2.8_{-0.1}^{+0.5}   $  &~~2.49187(7)$^{[3]}$               & $2.84(0.52) \times 10^{35}$   \\    
\vspace{1 mm}                                                                
32633~~         &214.5(2)        &10.2(0.4)~~~         &     $4.224(0.063)           $$^{[7]}$     & 2.9(0.2)           &2.6(0.4)            &$17_{-2}^{+1}        $  &~~6.4303(8)$^{[16]}$               & $3.85(1.21) \times 10^{35}$   \\    
\vspace{1 mm}                                                                
36526~~         &415.7(8)        &15(2)$^{[3]}$~~~~~\, &     $4.10(0.14)             $$^{[8]}$     & 2.35(0.1)          &4.3(0.2)$^{[8]}$    &$11(1)^{[8]}         $  &~~1.54185(4)$^{[17]}$              & $8.45(1.37) \times 10^{35} $  \\    
\vspace{1 mm}                                                                         
49333~~         &216(3)          &15.8(0.1)$^{[3]}$    &     $4.23(0.2)              $$^{[9]}$     & 3.6(0.2)           &4.5(0.2)            &$3.6_{-1.2}^{+1}     $  &~~2.1805(1)$^{[16]}$               & $2.95(2.02) \times 10^{35} $  \\    
\vspace{1 mm}                                                               
49976~~         &101.4(0.4)      &9.5(0.2)~~~          &     $4.298(0.063)           $$^{[7]}$     & 2.085(0.055)       &2.4(0.3)            &$7.5_{-1.7}^{+48.6}  $  &~~2.9768(2)$^{[16]}$               & $7.98(3.65) \times 10^{34} $  \\    
\vspace{1 mm}                                                               
55522~~         &277(3)          &16.2(0.2)~~~         &     $3.95(0.06)             $$^{[8]}$     & 4.8(0.2)           &5.2(0.3)            &$3.1(0.4)^{[8]}      $  &~~2.7292(3)$^{[17]}$               & $4.67(1.42) \times 10^{35}$   \\    
\vspace{1 mm}                                                               
62140~~         &94.4(0.2)       &7.8(0.1)~~~          &     $3.99(0.13)             $$^{[4]}$     & 2.32(0.06)         &1.9(0.3)            &$5.11_{-0.33}^{+1.05}$  &~~4.28677(3)$^{[18]}$               & $3.25(0.52) \times 10^{34}$   \\    
\vspace{1 mm}                                                               
65339~~         &97(2)           &8.3(0.1)~~~          &     $4.00(0.09)             $$^{[4]}$     & 2.41(0.08)         &2.0(0.3)            &$15.7_{-1.6}^{+2.3}  $  &~~7.993(2)$^{[16]}$                & $1.01(0.24) \times 10^{35} $  \\    
\vspace{1 mm}                                                               
74067~~         &76(1)           &10.6(0.1)~~~         &     $4.08(0.09)             $$^{[4]}$     & 2.21(0.05)         &2.7(0.4)            &$3.44_{-0.07}^{+12.2}$  &~~3.115(2)$^{[18]}$                 & $1.87(0.17) \times 10^{34}$   \\    
\vspace{1 mm}                                                               
103192~~        &95(5)$^{[2]}$   &11.8(0.4)$^{[4]}$    &     $3.83_{-0.11}^{+0.06}   $$^{[4]}$     & 3.9(0.4)           &3.6(0.3)$^{[4]}$    &$1.38_{-0.32}^{+1.08}$  &~~2.35666(2)$^{[19]}$              & $5.56(3.11) \times 10^{34}$   \\    
\vspace{1 mm}                                                               
120198~~        &92.8(0.3)       &10.4(0.1)~~~         &     $4.14_{-0.12}^{+0.11}   $$^{[4]}$     & 2.235(0.055)       &2.7(0.4)            &$1.6_{-0.36}^{+0.41} $  &~~1.3858(8)$^{[20]}$               & $2.16(0.99) \times 10^{34}$   \\    
\vspace{1 mm}                                                               
122532~~        &153.2(3)        &12.8(0.3)~~~         &     3.70(n.a.)$^{[10]}$                   & 2.8(0.1)           &3.5(0.2)            &$3.0_{-0.9}^{+0.7}   $  &~~3.6807(1)$^{[21]}$               & $2.55(1.58) \times 10^{34}$   \\    
\vspace{1 mm}                                                               
125248~~        &86(4)           &9.7(0.1)~~~          &     4.05(n.a.)$^{[10]}$                   & 1.86(0.06)         &2.5(0.3)            &$9.0_{-1.1}^{+1.3}   $  &~~9.302(1)$^{[16]}$                & $6.86(1.83) \times 10^{33} $  \\    
\vspace{1 mm}                                                               
130559~~        &75.0(0.5)       &9.3(0.7)~~~          &     $4.07(0.10)             $$^{[4]}$     & 2.1(0.2)           &1.71(0.09)          &$2.36_{-0.44}^{+0.77}$  &~~1.90798(1)$^{[18]}$               & $1.82(0.87) \times 10^{34}$   \\    
\vspace{1 mm}                                                               
133029~~        &156.5(0.6)      &12.3(0.2)~~~         &     $4.176(0.024)           $$^{[7]}$     & 2.6(0.1)           &3.5(0.2)            &$9.0_{-0.3}^{+5.0}   $  &~~2.8885(2)$^{[16]}$               & $2.58(0.45) \times 10^{35} $  \\    
\vspace{1 mm}                                                               
137909~~        &36(1)           &7.5(0.4)$^{[4]}$     &     $3.77_{-0.14}^{+0.11}   $$^{[4]}$     & 2.7(0.1)           &2.6(0.1)            &$5.2_{-1}^{+7.1}     $  &18.4877(15)$^{[22]}$               & $2.87(1.19) \times 10^{33}$   \\    
\vspace{1 mm}                                                              
140160~~        &67.8(0.3)       &9.3(0.1)~~~          &     $4.11_{-0.08}^{+0.06}   $$^{[4]}$     & 2.18(0.06)         &2.4(0.3)            &$1.18_{-0.26}^{+0.29}$  &~~1.5959(1)$^{[18]}$                & $8.38(3.77) \times 10^{33} $  \\    
\vspace{1 mm}                                                               
140728~~        &91.4(0.4)       &10.4(0.1)~~~         &     $4.06_{-0.12}^{+0.10}   $$^{[4]}$     & 2.42(0.06)         &2.7(0.35)           &$2.3_{-0.36}^{+0.46} $  &~~1.29559(2)$^{[18]}$               & $7.24(2.35) \times 10^{34} $  \\    
\vspace{1 mm}                                                               
142884~~        &173(1)          &10.2(0.5)~~~         &     $4.344(0.063)           $$^{[7]}$     & 2.82(0.09)         &2.6(0.4)            &$< 0.75              $  &~~0.80296(2)$^{[16]}$              & $4.05(0.92) \times 10^{34}$   \\    
\vspace{1 mm}                                                               
145102~~        &178(2)          &8.8(0.3)~~~          &     3.83(n.a.)$^{[11]}$                   & 3.8(0.1)           &2.2(0.3)            &$< 0.98              $  &~~1.4178(6)$^{[23]}$               & $9.43(2.12) \times 10^{34} $  \\    
\vspace{1 mm}                                                               
149822~~        &146(3)          &10.4(0.2)~~~         &     4.01(n.a.)$^{[11]}$                   & 2.75(0.08)         &2.7(0.3)            &$4.0_{-1.1}^{+2.1}   $  &~~1.966(1)$^{[3]}$                 & $1.72(0.96) \times 10^{35}$   \\    
\vspace{1 mm}                                                                
151346~~        &194(3)          &6.1(0.1)~~~          &     4.27(n.a.)$^{[11]}$                   & 4.8(0.25)          &2.1(0.1)            &$< 1.7               $  &~~2.180(2)$^{[23]}$                & $3.59(0.92) \times 10^{35}$   \\    
\vspace{1 mm}                                                              
152107~~        &53.2(0.3)       &8.9(0.1)~~~          &     $4.06_{-0.06}^{+0.05}   $$^{[4]}$     & 2.185(0.065)       &2.2(0.3)            &$4.2_{-0.2}^{0.5}    $  &~~3.8586(6)$^{[16]}$               & $1.89(0.26) \times 10^{34}$   \\    
\vspace{1 mm}                                                              
164224~~        &223(3)          &7.7(0.2)~~~          &     4.1577(n.a.)$^{[12]}$                 & 2.27(0.08)         &1.8(0.3)            &$\approx 1.7^{[13]}  $   &~~0.73(1)$^{[24]}$                 & $1.15(0.27) \times 10^{35} $  \\    
\vspace{1 mm}                                                              
168856~~        &202(1)          &8.3(0.4)~~~          &     $3.744(0.033)           $$^{[7]}$     & 3.6(0.2)           &3.15(0.2)           &$3.5_{-0.6}^{+3.7}   $  &~~2.4277(1)$^{[16]}$               & $2.80(1.06) \times 10^{35}$   \\    
\vspace{1 mm}                                                               
175132~~        &316(4)          &11.1(0.2)~~~         &     $3.858(0.02)            $$^{[7]}$     & 5.7(0.3)           &5.45(0.3)           &$> 3.5               $  &~~8.030(n.a.)$^{[3]}$              & $1.53(0.40) \times 10^{35}$   \\    
\vspace{1 mm}                                                               
183339~~        &320(5)          &14.0(0.1)~~~         &     3.87(n.a.)$^{[11]}$                   & 4.2(0.25)          &4.9(0.25)           &$> 4.5               $  &~~4.204(n.a.)$^{[3]}$              & $2.48(0.68) \times 10^{35}$   \\    
\vspace{1 mm}                                                              
224801~~        &178(1)          &12.3(0.2)~~~         &     3.7(n.a.)$^{[5]}$                     & 3.0(0.15)          &3.6(0.2)            &$> 4.6               $  &~~3.73980(n.a.)$^{[25]}$           & $8.27(2.11) \times 10^{34} $  \\    
\vspace{1 mm}                                                        
345439~~        &2330(140)       &14.5(0.3)~~~         &     $4.29(0.19)             $$^{[8]}$     & 3.7(0.5)$^{[8]}$   &8.3(0.8)$^{[8]}$    &$8.9(1.1)^{[8]}      $  &~~0.77018(2)$^{[26]}$              & $1.19(0.59) \times 10^{37}$   \\    
                                         

\hline
\end{tabular}
\vspace{-2. mm}
\begin{list}{}{}
\item[Notes:] 
$^{\ast}$\,For the star HD\,11502, we made a mistake when the present project was planned; we wrongly assumed the stellar rotation period of the close star HD\,11503 (separation about 7.5 arcsec). The following critical analysis of each target revealed our mistake; therefore, we cannot calculate the value of the $L_{\text{CBO}}$ of HD\,11502.
$^{\dag}$\,The uncertainty in the least significant digit is reported in parentheses.
\item[References:] 
$^{[1]}$\,\citet{Bailer-Jones2018};
$^{[2]}$\,\citet{hipparcos};
$^{[3]}$\,\citet{Shultz2022};
$^{[4]}$\,\cite{Sikora2019a};
$^{[5]}$\,\cite{Soubiran2016};
$^{[6]}$\,\cite{Glagolevskij2006};
$^{[7]}$\,\cite{Abdurrouf2022};
$^{[8]}$\,\cite{Shultz2019MNRAS490};
$^{[9]}$\,\cite{Cidale2007};
$^{[10]}$\,\cite{Ghazaryan2018};
$^{[11]}$\,\cite{Glagolevskij2019};
$^{[12]}$\,\cite{Fouesneau2022};
$^{[13]}$\,\cite{Bowman2018};
$^{[14]}$\,\cite{Renson&Catalano2001};
$^{[15]}$\,\cite{Musielok1980};
$^{[16]}$\,\cite{Bernhard2020};
$^{[17]}$\,\cite{Shultz2018_475};
$^{[18]}$\,\cite{Sikora2019b};
$^{[19]}$\,\cite{Manfroid&Renson1994};
$^{[20]}$\,\cite{Wade1998};
$^{[21]}$\,\cite{Lanz&Mathys1991};
$^{[22]}$\,\cite{Bagnulo1999};
$^{[23]}$\,\cite{Wraight2012};
$^{[24]}$\,\cite{Buysschaert2018};
$^{[25]}$\,\cite{CatalanoRenson1998}.
$^{[26]}$\,\cite{Hubrig2017}.

\end{list}

\end{center}

\end{table*}

\section{Details of the VLA observations}
\label{appendix:vla_obs}
The epochs when the observations were performed, the time on source, and the adopted calibrators for each target's observations are listed in Table\,\ref{tab:vla_obs}. In the table, the synthetic beams of the VLA interferometer, the noise measured in the calibrated maps in the clean region close to the sky position where the targets are located, with the notes reported by the VLA operators, are also reported.

\begin{table*}[h!]
\caption[ ]{Log of the VLA observations at the X-band ($\nu=9$ GHz; $\Delta \nu = 2$ GHz). Array Configuration: B. Code: 23A-002. Beam size and measured map noise are listed in the last three columns.}
\vspace{-4. mm}
\label{tab:vla_obs}
\footnotesize
\begin{center}
\begin{tabular}{@{}l@{~}c@{~}r            l@{~~}         l@{~}                 c          c                                     r           r}
\hline   
\hline   
Date-OBS        &UTC         &$\Delta$T   &Source        &Phase cal       &Flux cal  &FWHM                                 &PA         &RMS              \\
                &            &(min)       &              &                &          &($\prime\prime\times \prime\prime$)  &(degree)   &($\mu$Jy/beam)   \\
\hline      
2023-01-24$^a$  &14:52:34.5  &$16$        &HD\,137909    &J1513+2338      & 3c286    &$0.54\times0.63$                     &$64.12$    &16~~            \\
2023-01-24$^a$  &15:15:31.5  &$16$     	  &HD\,140160    &J1608+1029      & 3c286    &$0.55\times0.66$                     &$51.10$    &17~~             \\
2023-01-25$^b$  &12:58:13.5  &$16$        &HD\,125248    &J1432-1801      & 3c286    &$0.63\times1.04$                     &$4.02$     &13~~              \\
2023-01-25$^b$  &13:17:10.5  &$9$         &HD\,130559    &J1432-1801      & 3c286    &$0.66\times0.86$                     &$6.90$     &15~~             \\
2023-01-25$^b$  &13:30:37.5  &$9$         &HD\,122532    &J1352-4412      & 3c286    &$0.65\times3.06$                     &$13.65$    &21~~            \\                        
2023-02-01      &01:07:28.5  &$9$         &HD\,15089     &J0228+6721      & 3c147    &$0.56\times2.01$                     &$19.48$    &130$^{\dag}$\,           \\
2023-02-01      &01:20:25.5  &$9$         &HD\,21699     &J0349+4609      & 3c147    &$0.60\times0.67$                     &$15.14$    &10~~            \\
2023-02-01      &01:34:22.5  &$16$        &HD\,224801    &J2355+4950      & 3c147    &$0.63\times0.79$                     &$-79.56$   &7~~      \\                                  
2023-02-13      &12:07:19.5  &$16$        &HD\,142884    &J1626-2951      & 3c286    &$0.58\times1.37$                     &$-17.00$   &8~~      \\
2023-02-13      &12:26:46.5  &$16$        &HD\,145102    &J1626-2951      & 3c286    &$0.57\times1.52$                     &$-15.38$   &8~~      \\                                   
2023-02-14$^c$  &01:22:34.5  &$16$        &HD\,49976     &J0653-0625      & 3c147    &$0.63\times1.23$                     &$-35.32$   &12~~      \\
2023-02-14$^c$  &01:35:04.5  &$16$        &HD\,49333     &J0650-1637      & 3c147    &$0.59\times1.71$                     &$-31.57$   &13~~      \\                                 
2023-02-14$^c$  &06:56:37.5  &$16$        &HD\,55522     &J0648-3044      & 3c147    &$0.61\times1.94$                     &$28.29$    &9~~      \\
2023-02-14$^c$  &07:17:34.5  &$9$         &HD\,74067     &J0828-3731      & 3c147    &$0.59\times3.02$                     &$14.15$    &14~~      \\    
2023-02-16      &03:30:37.5  &$16$        &HD\,32633     &J0443+3441      & 3c147    &$0.61\times0.67$                     &$-14.42$   &7~~      \\
2023-02-16      &03:53:34.5  &$16$        &HD\,36526     &J0541-0541      & 3c147    &$0.66\times0.79$                     &$10.80$    &7~~      \\  
2023-02-16      &02:59:52.5  &$16$        &HD\,62140     &J0805+6144      & 3c147    &$0.59\times0.82$                     &$34.40$    &9~~      \\
2023-02-16      &03:18:49.5  &$9$         &HD\,65339     &J0805+6144      & 3c147    &$0.60\times0.81$                     &$34.95$    &12~~      \\  
2023-02-18      &00:18:34.5  &$9$         &HD\,12767     &J0153-3310      & 3c147    &$0.60\times1.52$                     &$11.61$    &11~~      \\
2023-02-18      &00:33:31.5  &$9$         &HD\,11502     &J0152+2207      & 3c147    &$0.64\times0.66$                     &$32.78$    &10~~      \\  
2023-03-15      &14:46:28.5  &$9$         &HD\,168856    &J1832-1035      & 3c286    &$0.61\times0.81$                     &$12.52$    &11~~      \\
2023-03-15      &14:59:25.5  &$16$        &HD\,164224    &J1820-2528      & 3c286    &$0.59\times1.22$                     &$19.51$    &13~~      \\ 
2023-03-17      &09:55:58.5  &$16$        &HD\,125248    &J1432-1801      & 3c286    &$0.63\times1.04$                     &$4.02$     &8~~      \\
2023-03-17      &10:13:55.5  &$9$         &HD\,130559    &J1432-1801      & 3c286    &$0.66\times0.86$                     &$6.90$     &10~~     \\ 
2023-04-05      &12:19:07.5  &$16$        &HD\,149822    &J1608+1029      & 3c286    &$0.63\times0.73$                     &$-87.50$   &8~~      \\
2023-04-05      &12:42:04.5  &$9$         &HD\,151346    &J1626-2951      & 3c286    &$0.68\times1.31$                     &$21.80$    &11~~      \\  
2023-04-11      &12:27:01.5  &$9$         &HD\,140728    &J1500+4751      & 3c286    &$0.59\times0.78$                     &$-71.61$   &10~~      \\
2023-04-11      &12:14:04.5  &$9$         &HD\,133029    &J1549+5038      & 3c286    &$0.60\times0.82$                     &$-83.63$   &10~~      \\   
2023-04-30      &02:56:04.5  &$16$        &HD\,120198    &J1419+5423      & 3c286    &$0.61\times0.82$                     &$74.67$    &8~~      \\
2023-04-30      &03:20:58.5  &$9$         &HD\,103192    &J1209-3214      & 3c286    &$0.59\times1.96$                     &$-14.34$   &11~~      \\
2023-04-30      &08:14:28.5  &$9$         &HD\,345439    &J1925+2106      & 3c286    &$0.63\times1.25$                     &$-65.60$   &13~~      \\
2023-04-30      &08:29:55.5  &$9$         &HD\,152107    &J1658+4737      & 3c286    &$0.59\times0.64$                     &$22.60$    &10~~      \\
2023-05-02      &08:29:37.5  &$16$        &HD\,175132    &J1845+4007      & 3c286    &$0.63\times0.74$                     &$77.58$    &8~~      \\
2023-05-02      &08:51:04.5  &$16$        &HD\,183339    &J1927+6117      & 3c286    &$0.61\times0.80$                     &$63.81$    &7~~      \\

\hline
\end{tabular}
\vspace{-2. mm}
\begin{list}{}{}
\item[Notes.] 
$^a$\,Snow accumulation on all dishes. Sensitivity may be degraded.
$^b$\,Snow on Primary estimated to be 15-30\% on ALL Antennas.
$^c$\,Approximately 5-10\% of dish surface covered in snow. 
$^{\dag}$\,Strong sidelobe contamination produced by the phase calibrator that is too close to the source sky position.
\end{list}
\end{center}

\end{table*}

\section{Proper motion correction}
\label{appendix:prop_motion}

The Earth proximity of the selected targets, combined with the high spatial resolution achieved by the VLA in B configuration at the observing frequency of 9\,GHz, could affect the effective sky position of the selected stars. This issue required the correction of the sources' positions on the sky at the observing epochs produced by their proper motions. The proper motions corrections were performed using the procedure {\sc calcpos} enabled within the {\sc idl} software package. The input J2000 targets' positions and the effective positions at the epochs when the VLA observations were performed are listed in Table\,\ref{tab:correct_pro_motion}. 

\begin{table*}[h!]
\caption[ ]{ICRS coordinates, correspondent proper motions (mostly from \citealp{hipparcos}), and radial velocities (mostly from \citealp{Kharchenko2007}) of the selected stars. The last two columns list the coordinates corrected for the proper motions reported at the observing epoch (year 2023).}
\vspace{-2. mm}
\label{tab:correct_pro_motion}
\footnotesize
\begin{tabular}{l            c                 c                                       r                    r                    r                   c                c                                     }
\hline      
\hline      
Source        &RA(J2000)        &DEC(J2000)                             &$\mu_{\text{RA}}$   &$\mu_{\text{DEC}}$  &RV                 &RA                &DEC                                   \\
              &(h:m:s)          &(${\circ}:{\prime}:{\prime\prime}$)    &(mas/yr)            &(mas/yr)            &(km/s)             &(h:m:s)           &(${\circ}:{\prime}:{\prime\prime}$)   \\
\hline      
HD\,11502     &01:53:31.819203  &$+19$:17:45.430794                     &$  77.845^{[1]} $   &$-106.965^{[1]} $   &$    3.7    $        &01:53:31.945669   &$+19$:17:42.970600                      \\
HD\,11503     &01:53:31.812928  &$+19$:17:37.879979                     &$  79.20        $   &$ -97.63        $   & $  -0.6    $        &01:53:31.941594   &$+19$:17:35.634486                      \\
HD\,12767     &02:04:29.445558  &$-29$:17:48.496909                     &$   1.849       $   &$  11.772       $   &$  18.5     $        &02:04:29.449285   &$-29$:17:48.226154                      \\
HD\,15089     &02:29:03.917010  &$+67$:24:08.592124                     &$ -12.304       $   &$   6.586       $   &$   1.2     $        &02:29:03.789237   &$+67$:24:08.743599                      \\
HD\,21699     &03:32:08.608420  &$+48$:01:24.528549                     &$  18.314       $   &$ -28.723       $   &$   1.8     $        &03:32:08.671196   &$+48$:01:23.867919                      \\
HD\,32633     &05:06:08.358482  &$+33$:55:07.279039                     &$   6.998       $   &$ -15.246       $   &$ -13.02    $        &05:06:08.374064   &$+33$:55:06.928381                      \\
HD\,36526     &05:32:13.081277  &$-01$:36:01.745708                     &$   0.465^{[2]} $   &$  -1.748^{[2]} $   &$ 30.2^{[3]}$        &05:32:13.081991   &$-01$:36:01.785911                      \\
HD\,49333     &06:47:01.483527  &$-21$:00:55.451696                     &$ -14.568       $   &$   5.778       $   &$  19.0     $        &06:47:01.457893   &$-21$:00:55.318803                      \\
HD\,49976     &06:50:42.303022  &$-08$:02:27.587803                     &$  -9.682       $   &$   2.626       $   &$  19.2     $        &06:50:42.287880   &$-08$:02:27.527405                      \\
HD\,55522     &07:12:12.214709  &$-25$:56:33.316083                     &$  -5.568       $   &$   9.286       $   &$  21.6     $        &07:12:12.204150   &$-25$:56:33.102505                      \\
HD\,62140     &07:46:27.414668  &$+62$:49:49.882911                     &$ -36.63        $   &$ -61.36        $   &$   1.9     $        &07:46:27.291667   &$+62$:49:48.471628                      \\
HD\,65339     &08:01:42.440387  &$+60$:19:27.806833                     &$ -25.627       $   &$ -26.731       $   &$  -2.2     $        &08:01:42.280075   &$+60$:19:27.192014                      \\
HD\,74067     &08:40:19.175746  &$-40$:15:49.936471                     &$ -57.208       $   &$   6.601       $   &$  21.1     $        &08:40:19.025101   &$-40$:15:49.784643                      \\
HD\,103192    &11:52:54.416874  &$-33$:54:29.332717                     &$ -56.56        $   &$  -0.19        $   &$-1.0       $        &11:52:54.520890   &$-33$:54:29.286720                      \\
HD\,120198    &13:46:35.656788  &$+54$:25:57.643559                     &$ -18.559       $   &$  -4.742       $   &$  -2.4     $        &13:46:35.572676   &$+54$:25:57.534491                      \\
HD\,122532    &14:03:27.467453  &$-41$:25:24.032197                     &$ -33.969       $   &$ -17.061       $   &$   4.0     $        &14:03:27.374817   &$-41$:25:24.424597                      \\
HD\,125248    &14:18:38.250561  &$-18$:42:57.465294                     &$ -65.51        $   &$ -36.08        $   &$ -10.9     $        &14:18:38.144504   &$-18$:42:58.295135                      \\
HD\,130559    &14:49:19.050841  &$-14$:08:56.492329                     &$ -72.6         $   &$ -26.2         $   &$  -4.1     $        &14:49:18.890236   &$-14$:08:57.309766                      \\
HD\,133029    &15:00:38.717898  &$+47$:16:38.791986                     &$  -8.999       $   &$  15.852       $   &$ -11.22    $        &15:00:38.687921   &$+47$:16:39.156583                      \\
HD\,137909    &15:27:49.731530  &$+29$:06:20.522375                     &$-180.17        $   &$  85.92        $   &$ -18.0     $        &15:27:49.415337   &$+29$:06:22.498535                      \\
HD\,140160    &15:41:47.414736  &$+12$:50:51.093706                     &$  39.045       $   &$ -2.923        $   &$   1.9     $        &15:41:47.477719   &$+12$:50:51.026476                      \\
HD\,140728    &15:42:50.760211  &$+52$:21:39.243860                     &$ -65.91        $   &$  30.10        $   &$ -16.1     $        &15:42:50.594720   &$+52$:21:39.936156                      \\
HD\,142884    &15:57:48.802145  &$-23$:31:38.294134                     &$  -8.040       $   &$ -17.283       $   &$  -4.1     $        &15:57:48.787481   &$-23$:31:38.691643                      \\
HD\,145102    &16:10:15.922123  &$-26$:54:32.879906                     &$ -18.794       $   &$ -32.305       $   &$   4.0     $        &16:10:15.885882   &$-26$:54:33.622920                      \\
HD\,149822    &16:36:42.938527  &$+15$:29:50.827224                     &$ -15.098       $   &$ -32.096       $   &$   0       $        &16:36:42.913596   &$+15$:29:50.089016                      \\
HD\,151346    &16:47:46.528540  &$-23$:58:27.525919                     &$ -10.862       $   &$ -19.528       $   &$ 24.0      $        &16:47:46.508591   &$-23$:58:27.975061                      \\
HD\,152107    &16:49:14.217098  &$+45$:58:59.950141                     &$  23.069       $   &$ -50.073       $   &$ -1.0      $        &16:49:14.290356   &$+45$:58:58.798460                      \\
HD\,164224    &18:00:57.615001  &$-20$:59:17.208524                     &$   2.800       $   &$ -11.983       $   &$0^{[4]}    $        &18:00:57.619926   &$-20$:59:17.484133                      \\
HD\,168856    &18:22:10.820549  &$-07$:29:55.602976                     &$  -8.049       $   &$ -16.004       $   &$ -9.8      $        &18:22:10.807993   &$-07$:29:55.971069                      \\
HD\,175132    &18:52:07.249932  &$+41$:22:59.680929                     &$  -3.764       $   &$  -1.112       $   &$-23.1      $        &18:52:07.239680   &$+41$:22:59.655353                      \\
HD\,183339    &19:25:46.764540  &$+58$:01:38.356203                     &$  -1.994       $   &$   4.560       $   &$-22.0      $        &19:25:46.753636   &$+58$:01:38.461084                      \\
HD\,224801    &00:00:43.634494  &$+45$:15:12.007060                     &$  17.745       $   &$   0.607       $   &$ -1.0      $        &00:00:43.689397   &$+45$:15:12.021020                      \\
HD\,345439    &19:58:48.489652  &$+23$:05:21.559730                     &$  -0.539^{[2]}$   &$  -5.168^{[2]}$   &$-33.8^{[3]}$        &19:58:48.488675   &$+23$:05:21.440866                      \\

\hline
\end{tabular}
\vspace{-2. mm}
\begin{list}{}{}
\item[References:] 
$^{[1]}$ \citet{gaia_dr2_2018}; 
$^{[2]}$ \citet{gaia2020}; 
$^{[3]}$ \citet{Jonsson2020}; 
$^{[4]}$ \citet{Kervella2022}.

\end{list}

\end{table*}

\begin{figure}[h!]
\centering
\includegraphics[width=.94\columnwidth]{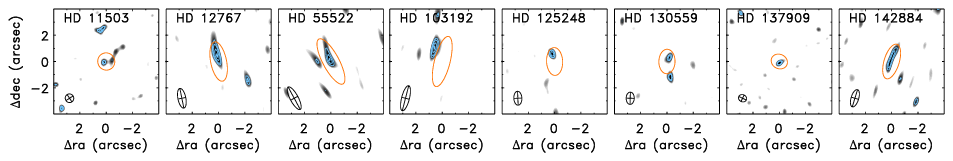}
\includegraphics[width=.94\columnwidth]{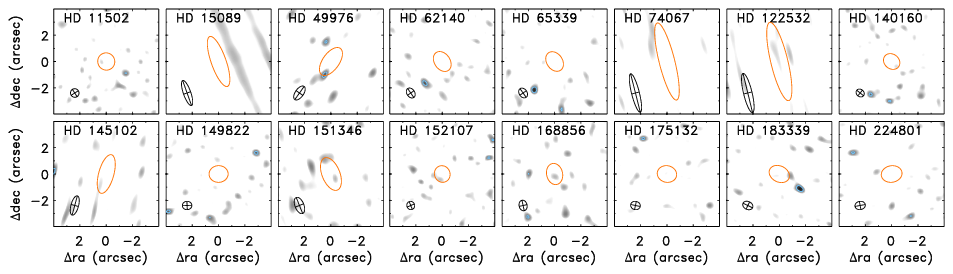}
\vspace{-4. mm}
\caption{
Top panels: maps of the 8 early-type magnetic stars tentatively detected at 9 GHz (same caption as Fig.\,\ref{fig:maps_detected}). 
Middle and bottom panels: maps of the 16 early-type magnetic stars not detected at 9 GHz, grey levels display the pixels map above the $1.5\sigma$ threshold. The $3\sigma$ level is marked by the light-blue contours. No reliable detection falls within the sky area covered by the FNBW (red ellipses) at the positions where the targets are expected.
}
\label{fig:maps_tentativ_undetected}
\end{figure}

The cleaned maps centered at the corrected sources' positions have been produced. Of the 32 targets, we confidently detect radio emission from 9 targets($\approx 28\%$) , and tentatively from another 8 targets (25\%). 
The maps of the 9 sources showing a robust detection of their radio emission are displayed in Fig.\,\ref{fig:maps_detected} of the main text. The maps of the sources having tentative detections (that are those sources having peak intensity higher than $3\sigma$ but lower than the threshold of  $5\sigma$) and the remaining non-detected sources are displayed in Fig.\,\ref{fig:maps_tentativ_undetected}.

\clearpage
\section{The simulated spectra}
\label{appendix:ref_star_spectra}

\begin{figure*}[h!]
\centering
\includegraphics[scale=.8]{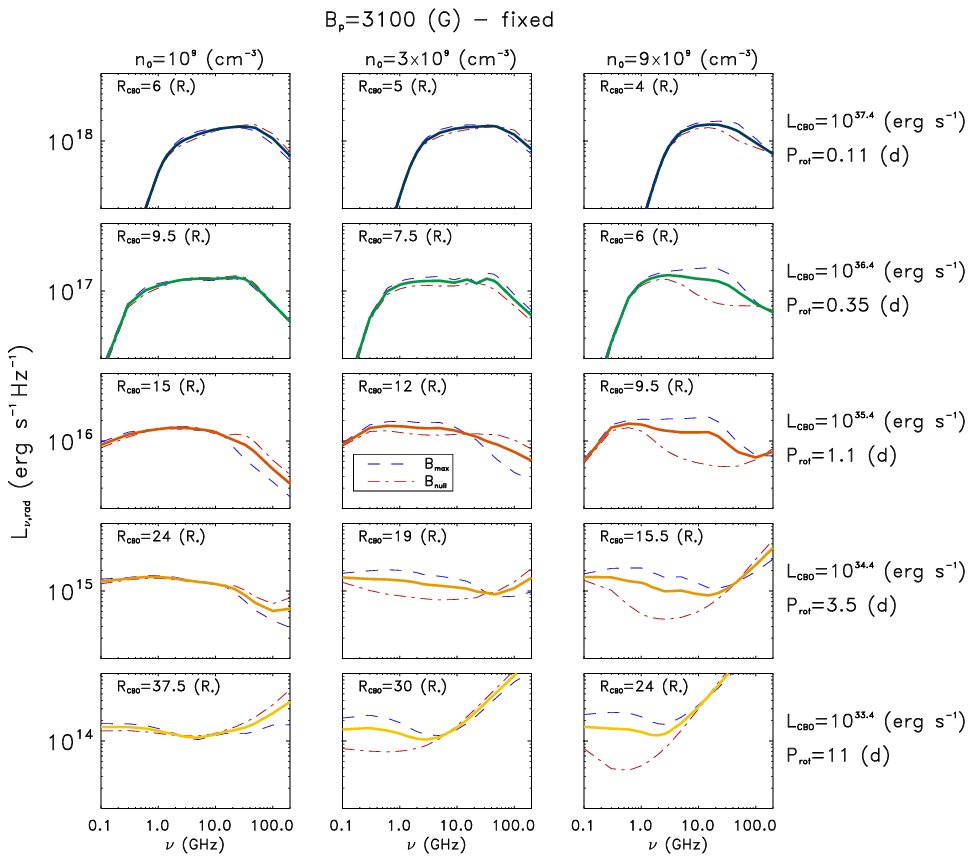}
\caption{Simulated spectra calculated taking the magnetic field fixed: $B_{\text{p}}=3100$ G, and varying the rotation period ($P_{\text{rot}}$) to cover the range of the CBO powers of the stars displayed in Fig.\,\ref{fig:cbo_vs_radio}, in accordace with Eq.\,\ref{eq:lrad_lcbo}. Blue dashed lines refer to calculations performed at the rotational phase when the magnetic pole is better visible ($B_{\mathrm {max}}$). Red dot-dashed lines refer to the calculations performed when the dipole axis is perpendicularly oriented with respect to the line of sight; in this case, the effective magnetic field is null ($B_{\mathrm {null}}$). The thick line represents the average spectrum between the two, calculated using the different magnetospheric orientations. The several $L_{\text{CBO}}$ analyzed have been marked using the same color code adopted in Fig.\,\ref{fig:nrl_per_l}.
}
\label{fig:spe_maps_simul_Bfixed}
\end{figure*}

\begin{figure*}[h!]
\centering
\includegraphics[scale=.8]{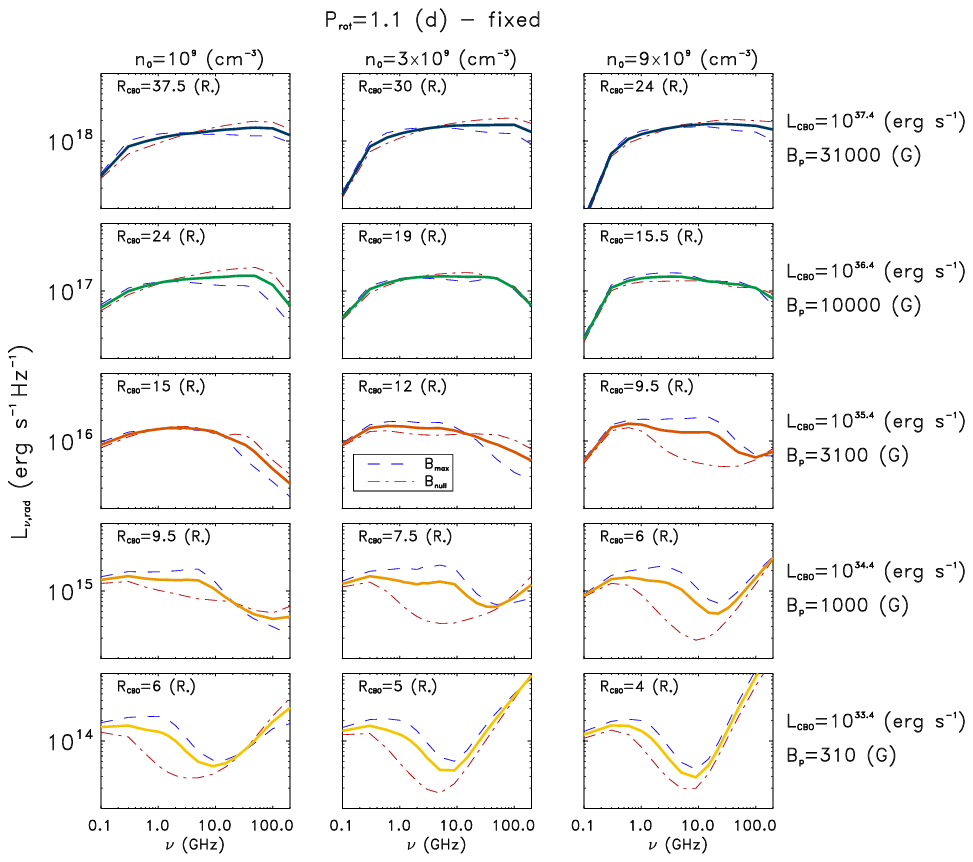}
\caption{Same caption of Fig.\,\ref{fig:spe_maps_simul_Bfixed}. In this case, the rotation period has been assumed fixed: $P_{\text{rot}}=1.1$ d, whereas $L_{\text{CBO}}$ has been changed by varying the polar field strength ($B_{\text{p}}$).
}
\label{fig:spe_maps_simul_Pfixed}
\end{figure*}

The main result of the radio spectra modeling exercise described in Sec.\,\ref{sec:spectrum_calculation} was to verify that the column density of the relativistic electrons is related to the power released by the centrifugal breakouts, as discussed in Sec.\,\ref{sec:cbo_effic}. As a secondary result, the radio spectra calculations also evidenced that the different model parameters, listed in Table\,\ref{tab:param_simul}, adopted to cover the range of the measured radio luminosities of the BA-type magnetic stars distributed in the $L_{\nu, {\mathrm {rad}}}$--$L_{\mathrm{CBO}}$ diagram (Fig.\ref{fig:cbo_vs_radio}), produce calculated radio spectra characterized by different shapes. The calculated spectra are shown in Figs.\,\ref{fig:spe_maps_simul_Bfixed}~and\,\ref{fig:spe_maps_simul_Pfixed}. 

In the following, we provide a qualitative explanation of the different shapes of the simulated spectra depending on the different model parameters adopted.

Looking at Figs.\,\ref{fig:spe_maps_simul_Bfixed}~and\,\ref{fig:spe_maps_simul_Pfixed} two overal spectrum behaviour are evident. At the higher radio luminosities, the radio spectra evidenced three clear spectral regimes: rising, flat, and decaying regions. At the lower luminosities, the simulated spectra instead evidenced the clear inversion of the decaying spectral trend as the radio frequency increases  (panels pictured in the two bottom lines of Figs.\,\ref{fig:spe_maps_simul_Bfixed}~and\,\ref{fig:spe_maps_simul_Pfixed}). Such behavior is simply explained since when the non-thermal radio emission is low, the contribution provided by the thermal radio emission component due to the thermal plasma trapped within the corotating magnetosphere, mainly the equatorial {plasma disk}, becomes non-negligible. In the optically thick regime, the thermal radio emission has a positive spectral index, making this thermal component better visible at the higher frequencies. Furthermore, the viewing angle of the equatorial {plasma disk} produces a strong spectral difference between the two radio spectra calculated when the line of sight lies in the equatorial plane ($B_{\text{null}}$), and when this plane is strongly inclined ($B_{\text{max}}$). For the adopted oblique rotator model geometry, this inclination angle is $45^{\circ}$. In general, these simulated effects become more evident as the electron density of the equatorial {plasma disk} increases.

At the higher radio luminosities, the non-thermal radio component dominates the thermal one at almost all the simulated radio frequencies and, as general behaviour, the radio emission calculated when the magnetic pole is better visible ($B_{\text{max}}$) is higher than the corresponding radio emission calculated when the magnetic dipole axis is perpendicular to the line of sight ($B_{\text{null}}$). Such simulated behaviour is well in accordance with the observed one, as examples see the cases of HD\,182180 \citep{Leto2017} or HD\,142184 \citep{Leto2018}. Even if, in several cases, it was simulated that above a certain frequency, such behaviour inverts, the star becomes brighter when the rotational phase coincides with the magnetic null.

\begin{table*}[h!]
\caption[ ]{Additional informations to those reported in Table~\ref{tab:param_simul}. The expected fluxes from the reference star located at 125\,pc are listed in column 3. Columns 5, 7, and 9 list the magnetic field strength at the distances where the CBO occurs.
}
\vspace{-5. mm}
\label{tab:param_simul_appendix}
\footnotesize
\begin{center}
\begin{tabular}{c c c r cc r cc r cc}                         
\hline                           
\hline    
\multicolumn{12}{l}{$B_{\text{p}}=3100$\,(G)}       \\
\hline    

~ &~ &~ &~  &\multicolumn{2}{c}{$n_0=10^9$\,(cm$^{-3}$)} &~ &\multicolumn{2}{c}{$n_0=3\times 10^9$\,(cm$^{-3}$)} &~ &\multicolumn{2}{c}{$n_0=9\times 10^9$\,(cm$^{-3}$)} \\
\cline{5-6}
\cline{8-9}
\cline{11-12}
$P_{\text{rot}}$ &$L_{\nu,\text{rad}}$        &$S_{\text I}$@125\,pc    &~ &$R_{\text{CBO}}$  &$B_{\text{CBO}}$     &~ &$R_{\text{CBO}}$  &$B_{\text{CBO}}$     &~ &$R_{\text{CBO}}$   &$B_{\text{CBO}}$      \\
(d)              &(erg\,s$^{-1}$\,Hz$^{-1}$)  &                         &~ &(R$_{\ast}$)      &(G)                  &~ &(R$_{\ast}$)      &(G)                  &~ &(R$_{\ast}$)       &(G)                   \\

\cline{5-6}
\cline{8-9}
\cline{11-12}

\vspace{-2 mm}                                                                    
 &&&&&&&&&&& \\

0.11            &$\approx 10^{18.2}$          &$\approx 85$\,mJy~~~~~   &~ &$\approx 6~~~$    &$\approx 43.1$~~     &~ &$\approx 5~~~$    &$\approx 62.0$~~    &~ &$\approx 4~~~$     &$\approx 96.9$~~           \\
0.35            &$\approx 10^{17.2}$          &$\approx 8$\,mJy~~~      &~ &$\approx 9.5$     &$\approx 17.2$~~     &~ &$\approx 7.5$     &$\approx 27.6$~~    &~ &$\approx 6~~~$     &$\approx 43.1$~~           \\
1.1~~           &$\approx 10^{16.2}$          &$\approx 0.8$\,mJy       &~ &$\approx 15~~~~~$ &$\approx 6.9$        &~ &$\approx 12~~~~~$ &$\approx 10.8$~~    &~ &$\approx 9.5$      &$\approx 17.2$~~         \\
3.5~~           &$\approx 10^{15.2}$          &$\approx 85$\,$\mu$Jy~~  &~ &$\approx 24~~~~~$ &$\approx 2.7$        &~ &$\approx 19~~~~~$ &$\approx 4.3$       &~ &$\approx 15.5~~$   &$\approx 6.5$          \\
11~~~~~~~       &$\approx 10^{14.2}$          &$\approx 8$\,$\mu$Jy     &~ &$\approx 37.5~~$  &$\approx 1.1$        &~ &$\approx 30~~~~~$ &$\approx 1.7$       &~ &$\approx 24~~~~~$  &$\approx 2.7$          \\


\hline   

\multicolumn{12}{l}{$P_{\text{rot}}=1.1$\,(d)}       \\
\hline

~ &~ &~ &~  &\multicolumn{2}{c}{$n_0=10^9$\,(cm$^{-3}$)} &~ &\multicolumn{2}{c}{$n_0=3\times 10^9$\,(cm$^{-3}$)} &~ &\multicolumn{2}{c}{$n_0=9\times 10^9$\,(cm$^{-3}$)} \\
\cline{5-6}
\cline{8-9}
\cline{11-12}
$B_{\text{p}}$ &$L_{\nu,\text{rad}}$         &$S_{\text I}$@125\,pc   &~ &$R_{\text{CBO}}$  &$B_{\text{CBO}}$     &~ &$R_{\text{CBO}}$    &$B_{\text{CBO}}$ &~ &$R_{\text{CBO}}$ &$B_{\text{CBO}}$   \\
(G)            &(erg\,s$^{-1}$\,Hz$^{-1}$)   &                        &~ &(R$_{\ast}$)      &(G)                  &~ &(R$_{\ast}$)        &(G)              &~ &(R$_{\ast}$)     &(G)                \\

\cline{5-6}
\cline{8-9}
\cline{11-12}

\vspace{-2 mm}                                                                    
 &&&&&&&&&&& \\

31000          &$\approx 10^{18.2}$          &$\approx 85$\,mJy~~~~~  &~ &$\approx 37.5~~$  &$\approx 11.0~~$     &~ &$\approx 30~~~~~$   &$\approx 17.2~~$    &~ &$\approx 24~~~~~$   &$\approx 26.9~~$   \\
10000          &$\approx 10^{17.2}$          &$\approx 8$\,mJy~~~     &~ &$\approx 24~~~~~$ &$\approx 8.7$        &~ &$\approx 19~~~~~$   &$\approx 13.8~~$    &~ &$\approx 15.5~~$    &$\approx 20.8~~$   \\
~~3100    &\multicolumn{3}{c}{Parameters already listed}  &-- &-- & &-- &-- & &-- &--  \\
~~1000         &$\approx 10^{15.2}$          &$\approx 85$\,$\mu$Jy~~ &~ &$\approx 9.5$     &$\approx 5.5$        &~ &$\approx 7.5$       &$\approx 8.9$        &~ &$\approx 6~~~$      &$\approx 13.9~~$   \\
~~~~310        &$\approx 10^{14.2}$          &$\approx 8$\,$\mu$Jy    &~ &$\approx 6~~~$    &$\approx 4.3$        &~ &$\approx 5~~~$      &$\approx 6.2$        &~ &$\approx 4~~~$      &$\approx 9.7$      \\


\hline    



\end{tabular}

\end{center}

\end{table*}

To cover the observed range of $L_{\text {CBO}}$, we separately varied two stellar parameters: the rotation period and the polar magnetic field strength. The simulated radio spectra calculated assuming the stellar magnetic field fixed ($B_{\text p}=3100$\,G, shown in Fig.\,\ref{fig:spe_maps_simul_Bfixed}) evidence that the low-frequency turnover moves at higher frequencies as the rotation period decreases (and consequently the region where the CBO occurs moves closer to the stellar surface). Such spectral displacement of the low-frequency turnover is not observed in the spectra calculated assuming the rotation period fixed ($P_{\text {rot}}=1.1$\,d, shown in Fig.\,\ref{fig:spe_maps_simul_Pfixed}), remaining almost fixed close to 200\,MHz.

The simulated spectral behavior described above can be explained by the strength of the magnetic field in the region where the non-thermal electrons are injected within the stellar magnetosphere. The radial distance where the non-thermal electrons are injected ($R_{\text{CBO}}$) is univocally defined once the stellar parameters are assigned. In the simulations, the adopted $R_{\text{CBO}}$ value is approximated to within half a stellar radius of the value numerically estimated using Eq.\,\ref{eq:r_cbo}. In the Table\,\ref{tab:param_simul_appendix}, the corresponding local values of the magnetic field strength ($B_{\text{CBO}}$) following the equatorial radial dependence described by a monopole field, calculated using Eq.\,\ref{eq:campo_mag_monopolo}, are also listed. We noticed that the $B_{\text{CBO}}$ values changed significantly in cases when $B_{\text p}$ was left fixed. On the contrary, the local magnetic field strength of the CBO region does not change strongly in cases when the $P_{\text {rot}}$ is fixed.

The results of the exercise discussed above suggest the need for further highly sensitive radio observations. In fact, many theoretical predictions of spectral behavior occur in magnetic BA stars with low radio luminosity. Their expected fluxes among the wide spectral range analyzed are very low, making their study a difficult challenge for the present state-of-the-art radio interferometers. On the other hand, this will be an optimal science case for the forthcoming new generation radio observing facilities (mostly SKA and ngVLA, which will offer high-sensitivity radio observations covering an ultra-wide spectral range).

\end{appendix}


\end{document}